\documentclass{jpp}
\usepackage{graphicx}

\usepackage[utf8]{inputenc}
\usepackage[T1]{fontenc}
\usepackage{amsmath}
\newcommand{\deri}[2]{\frac{\partial #1}{\partial #2}}
\newcommand{\diff}[2]{\frac{1}{r}\frac{\partial}{\partial r}r#1 \frac{\partial #2}{\partial r}}
\renewcommand{\vec}[1]{\boldsymbol{#1}}
\newcommand{\pmax}{p_{\text{max}}}
\newcommand{\nRE}{n_{\text{RE}}}

\shorttitle{Runaway electron transport due to magnetic perturbations}
\shortauthor{P. Svensson, O. Embreus, S. L. Newton, K. Särkimäki, O. Vallhagen and T. Fülöp}

\title{Effects of magnetic perturbations and radiation on the runaway avalanche}

\author{P. Svensson\aff{1},
  O. Embreus\aff{1}, 
  S. L. Newton\aff{2}, 
 K. Särkimäki\aff{1}, O.~Vallhagen\aff{1}
 \and  T. Fülöp\aff{1}  \corresp{\email{tunde@chalmers.se}}}

\affiliation{\aff{1}Department of Physics, Chalmers University of Technology, SE-41296 Göteborg, Sweden
\aff{2}CCFE, Culham Science Centre, Abingdon, Oxon OX14 3DB, UK}

\usepackage[textsize=tiny]{todonotes}

\usepackage{pgfplots}
\pgfplotsset{compat=1.3}
\pgfplotsset{plot coordinates/math parser=false}
\newlength\figureheight
\newlength\figurewidth
\usepackage{subcaption}
\usepackage{float}

\newcommand{\dt}[1]{\deri{#1}{t}}
\newcommand{\derip}[1]{\deri{#1}{p}}

\newcommand{\loglambda}{\ln \Lambda}
\newcommand{\pperp}{p_{\perp}}
\newcommand{\ppar}{p_\|}

\newcommand{\Eceff}{E_c^{\text{eff}}}
\renewcommand{\vec}[1]{\boldsymbol{#1}}

\begin{document}

\maketitle

\begin{abstract}
    The electron runaway phenomenon in plasmas depends sensitively on
    the momentum-space dynamics.  However, efficient simulation of the
    global evolution of systems involving runaway electrons typically
    requires a reduced fluid description.  This is needed for example
    in the design of essential runaway mitigation methods for
    tokamaks.  In this paper, we present a method to include the
    effect of momentum-dependent spatial transport in the runaway
    avalanche growth rate. We quantify the reduction of the growth
    rate in the presence of electron diffusion in stochastic magnetic
    fields and show that the spatial transport can raise the effective
    critical electric field. Using a perturbative approach we derive a
    set of equations that allows treatment of the effect of spatial
    transport on runaway dynamics in the presence of radial variation
    in plasma parameters. This is then used to demonstrate the effect
    of spatial transport in current quench simulations for ITER-like
    plasmas with massive material injection. We find that in scenarios
    with sufficiently slow current quench, due to moderate impurity and deuterium injection, the
    presence of magnetic perturbations reduces the final runaway current
    considerably. Perturbations localized at the edge are not
    effective in suppressing the runaways, unless the runaway
    generation is off-axis, in which case they may lead to formation
    of strong current sheets at the interface of the confined and
    perturbed regions.
\end{abstract}

\section{Introduction}
Electron runaway is seen as one of the main threats to successful operation of magnetic confinement fusion devices with large plasma currents, such as ITER \citep{Lehnen2015, Breizman_2019}. The number of e-foldings in the runaway avalanche during a plasma-terminating disruption increases drastically when a tokamak is scaled up to ITER parameters from those currently in operation \citep{RosenbluthPutvinski1997}. This calls for accurate models for the runaway generation and losses to ensure the design of a successful disruption mitigation system \citep{HollmannDMS}. 
    
There is a wealth of experimental evidence that magnetic perturbations, occurring  either naturally after a disruption or induced by external magnetic coils, can prevent or reduce runaway electron beam formation. In JET, a high level of magnetic fluctuations following a disruption has been seen to correlate with the absence of runaways \citep{Gill_2002}.  Broadband magnetic turbulence has been observed to lead to suppression of runaway current if the perturbation exceeds a certain level also in TEXTOR \citep{Zeng2013} and in J-TEXT \citep{Zeng2017}. Kinetic instabilities driven by the runaways themselves can also induce local magnetic perturbations increasing the radial transport. Observations at DIII-D indicate that when the power in the instabilities exceeds a threshold, runaway plateau formation is absent \citep{Lvovskiy2018}. Perturbations imposed by external magnetic coils have also been shown to suppress the formation of runaway beams in several tokamaks \citep{Yoshino_2000,Lehnen_2008,Lehnen2009,Mlynar_2018}.

The avalanche generation of runaway electrons is a result of momentum transfer between an existing runaway electron and a thermal one in a close collision. This leads to a growth of the runaway population that is proportional to the existing number of runaway electrons. Consequently, as the radial transport of runaways is also proportional to their number, it can reduce the growth rate of the exponentiation \citep{helander2000suppression}. Perturbations in the plasma confining magnetic field result in spatial transport and subsequent losses of runaway electrons \citep{rechester1978electron}. These losses reduce the number of runaway electrons participating in the avalanche mechanics and thereby have the potential to reduce the conversion of the initial plasma current to a runaway beam.  

Modelling of a disrupting tokamak plasma resolved both in momentum, as needed for the runaway problem, and spatially, as needed to describe the evolution of plasma parameters, is computationally costly. Therefore, to follow the evolution of the disruption, simplified fluid models for the runaway populations are often used \citep{smith2006runaway,Papp_2013,Matsuyama_2017,Bandaru2019,fulop2020effect}. In these, the momentum space dynamics has been captured approximately, and an effective theory only dependent on spatially varying quantities is used to describe the growth and loss of the runaway population. Such simplified disruption modelling has been used to estimate the post-disruption runaway population in the presence of massive material injection \citep{martin2017formation,Vallhagen-JPP-20,Linder_2020}. However, these studies focused on the generation rates of the runaways and neglected the losses due to spatial transport.

The transport due to the perturbations is in general momentum dependent \citep{hauff2009runaway}, preventing a straightforward fluid description of the phenomena. The goal of this paper is to present a theory describing an effective rate of generation for runaway electrons which incorporates the effects of a momentum-dependent spatial diffusion. The diffusion considered here may originate from the motion of electrons in regions of stochastic magnetic fields as well as other perturbed magnetic field structures. In Sec. \ref{sec:homogenous} we derive a self-consistent expression for the reduced avalanche growth rate of runaway electrons, including the effect of spatial transport, as well as radiation reaction forces and partially ionised impurities in a homogeneous plasma. Spatial variations in the plasma are investigated in Sec.~\ref{sec:inhomogeneous} with a perturbation approach which conserves particle number to investigate the impact of radial transport in more realistic disruption simulations. 

We find that, if the time-scale of the losses is comparable with that of the avalanche, spatial transport can raise the critical electric field for runaway generation significantly. The reason is that, even as runaway electrons are generated through close collisions by the avalanche dynamics, there need not be a net growth of the population if the relativistic electrons are transported out of the plasma.  This may be part of the explanation of experimental observations which show strongly elevated critical electric fields \citep{MartinSolis,Hollmann2013,Granetz,Paz-Soldan,Popovic}. The value of the critical field is also important for the dynamics in the current decay phase of disruptions, where the electric field tends to a value at which the loss and gain of runaway electrons is balanced \citep{BreizmanAleynikov2017Review}. 

We demonstrate the effect of magnetic perturbations on runaway evolution in simplified disruption simulations in Sec.~\ref{sec:iter}, taking into account the evolution and transport of runaways self-consistently with the electric field. We consider ITER-like plasmas with a combination of neon and deuterium injection and find that the runaway current can be suppressed, if the perturbations reach all the way to the plasma centre.   The mixed magnetic topology common in disruptions is seen to have the potential to generate strong current sheets. Their stability may in turn be expected to impact the magnetic perturbation profile.  \section{Radial diffusion of runaway electrons in the presence of radiation}
   Runaway electrons are almost collisionless and, as such, tend to follow magnetic field lines closely. Thus, in a stochastic magnetic field, the trajectories of runaway electrons generated close to one another will diverge with a rate dependent on the particle velocity along the field line and the rate of divergence of nearby field lines themselves \citep{rechester1978electron}. For a population of runaway electrons this leads to diffusive cross-field transport, the magnitude of which depends on the perturbation strength. However, at relativistic energies the electrons do not follow field lines closely, which causes the transport to decrease with increasing energy due to the effects associated with the finite orbit width \citep{Myra_1992,hauff2009runaway,Sarkimaki_2020}. 
   Furthermore, it has been shown that modelling the transport as purely diffusive is insufficient in mixed magnetic topologies containing both islands and stochastic regions \citep{Papp_2015}, but this can be addressed by including an advection term in the model \citep{Sarkimaki_2016}.
   A simplified theory to account for the momentum dependent radial diffusion in the avalanche growth rate was proposed by \cite{helander2000suppression}, a theory which we will build on and extend to account for radiation reaction forces and the presence of partially ionised impurities. We address a case with mixed magnetic topologies in Sec.~\ref{sec:iter}.

   The momentum-space dynamics in the electron runaway problem is often described by the high-energy limit of the gyro-averaged kinetic equation with an accelerating electric field parallel to the magnetic field $\vec{B}$ 
   \citep{hesslow2018effect}:
    \begin{equation}
        \begin{aligned}
            \deri{f}{t} + \frac{ E}{\tau} \left(\xi\deri{f}{p} + \frac{1 - \xi^2}{p}\deri{f}{\xi} \right) =  C\{f\}-  \deri{}{\vec{p}}\cdot\left( \vec{F}_{\text{rad}}f\right).
        \end{aligned}
        \label{eq:full_kinetic_equation}
    \end{equation}
    Here, $f$ is the electron distribution function,  $p$ is the momentum normalised to $m_e c$, $\xi = \vec{p}\cdot\vec{B}/(pB)$ is the cosine of the pitch angle, $E$ is the electric field strength normalised to the critical electric field  $
        {E_c = n_e e^3 \ln{\Lambda_c} / \left(4 \pi \varepsilon_0^2 m_e c^2 \right)}$ \citep{connor1975relativistic},
  where $n_e$ is the electron density, $e$ the elementary charge, $m_e$ the electron mass, $\varepsilon_0$ the permittivity of free space,  $c$ the speed of light and $\ln{\Lambda_c} \simeq 14.6 +0.5 \ln{T_{\rm eV}/n_{\rm e20}}$ is the relativistic Coulomb logarithm, with $T_{\rm eV}$ being the temperature measured in electronvolts and $n_{\rm e20}$ the electron density normalised to $10^{20}\;\rm m^{-3}$. The relativistic collision time between electrons is ${\tau = m_e c / \left(e E_c\right)}$,  $C\{ f \}$ is the relativistic collision operator and the last term on the right hand side of \eqref{eq:full_kinetic_equation} represents radiation reaction forces from synchrotron radiation and bremsstrahlung.
    
    The relativistic test-particle collision operator is given by  \citep{helander2005collisional}
        \begin{equation}
            C\{ f \} = \nu_D \frac{1}{2}\deri{}{\xi}\left(1 - \xi^2\right)\deri{}{\xi}f + \frac{1}{p^2}\deri{}{p} \left(p^3 \nu_s f\right),
            \label{eq:basic_rel_coll_operator}
        \end{equation}
    where $\nu_s$ and  $\nu_D$ are the slowing down and deflection frequencies, respectively. For relativistic electrons $\nu_s$ and $\nu_D$ take the form 
    \begin{equation}
        \nu_D = \tau^{-1} \frac{\gamma}{p^3}\Bar{\nu}_D, \hspace{1cm} \nu_s = \tau^{-1} \frac{\gamma^2}{p^3}\Bar{\nu}_s,
    \end{equation}
    where for the case of a fully ionised plasma $\Bar{\nu}_s = 1$ and $\Bar{\nu}_D = 1 + Z_{\text{eff}}$ \citep{hesslow2018effect}. Here $\gamma=\sqrt{1+p^2}$ is the Lorentz factor, $Z_{\text{eff}} = n_e^{-1}\sum_{j} n_j Z_j^2$ is the effective charge and $j$ is an index which runs over all ion species in the plasma, each with density $n_j$ and charge $Z_j$.  
    
     In partially ionised plasmas, the slowing-down and deflection frequencies are
influenced by the extent to which fast electrons can penetrate the bound electron cloud
around the impurity ion, i.e. the effect of partial screening \citep{martinsolis1, BreizmanAleynikov2017Review}. The collision frequencies $\nu_s$ and $\nu_D$ can be generalised to account for the differences in the collisional dynamics at different energy scales arising when screening effects are introduced, and in the relativistic limit take the form \citep{hesslow2017effect,HesslowJPP}
    \begin{subequations}
        \begin{align}
            \Bar{\nu}_D &\approx 1 + Z_{\text{eff}} + \frac{1}{\ln{\Lambda_c}}\sum_{j} \frac{n_j}{n_e}\left[\left(Z_j^2 - Z_{0j}^2\right)\ln{\Bar{a}_j} - \frac{2}{3}N_{j}^2\right] + \frac{\ln{p}}{\ln{\Lambda_c}}\sum_{j}\frac{n_j}{n_e}Z_j^2,\\
            \Bar{\nu}_s &\approx 1 + \frac{1}{\ln{\Lambda_c}}\sum_{j} \frac{n_j}{n_e}N_{j}\left( \ln{I_j^{-1}} - 1 \right) + \frac{\ln{p}}{2\ln{\Lambda_c}}\left(1 + 3\sum_{j}\frac{n_j}{n_e}N_{j} \right),
        \end{align}
    \end{subequations}
    which we will denote as $\Bar{\nu}_D \approx \Bar{\nu}_{D0} + \Bar{\nu}_{D1}\ln{p}$ and $\Bar{\nu}_s = \Bar{\nu}_{s0} + \Bar{\nu}_{s1} \ln{p}$. The collision frequencies now depend on atomic parameters of species $j$: ionisation degree $Z_{0j}$,  charge number $Z_j$, number of bound electrons of the nucleus for species $j$, $N_j = Z_j - Z_{0j}$, mean excitation energy of the ion $I_j$  and  effective ion size $\Bar{a}_j$ determined from density functional theory calculations, given by \cite{hesslow2017effect}.  
     The effect of partially ionised ions in the plasma will influence the runaway generation \citep{hesslow2019influence,hesslow2019evaluation} as well as increase the critical electric field \citep{hesslow2018effect}.

    Synchrotron radiation and bremsstrahlung hinder the acceleration of runaway electrons. The effective term in the kinetic equation resulting from synchrotron radiation is \citep{Stahl2015,hirvijoki2015guiding,hirvijoki2015radiation}
    \begin{equation}
        \deri{}{\vec{p}}\cdot\left(\vec{F}_{\text{syn}} f\right) = -\frac{1}{p^2}\deri{}{p}\left( \frac{p^3\gamma}{\tau_{\text{syn}}}\left(1 - \xi^2\right)f\right) + \deri{}{\xi}\left( \frac{\xi \left( 1 - \xi^2 \right)}{\tau_{\text{syn}}\gamma}f\right),
    \end{equation}
    where $\tau_{\text{syn}}$ is the synchrotron radiation-damping time scale 
    \begin{equation}
        \tau_{\text{syn}} = 6\pi \varepsilon_0 m_e^3 c^3 / \left( e^4 B^2\right).
    \end{equation}
  Similarly to the treatment by \cite{hesslow2017effect}, the effect of bremsstrahlung is here incorporated into the kinetic equation via a mean-force model, which has been shown to capture the mean-energy accurately \citep{embreus2016effect},
    \begin{equation}
        \deri{}{\vec{p}}\cdot\left( \vec{F}_{\text{br}}f\right) = -\frac{1}{p^2}\deri{}{p}\left(p^2 F_{\text{br}}f\right).
    \end{equation}
Here $F_{\text{br}}$ is approximated by
    \begin{equation}
        F_{\text{br}} \approx \frac{p \alpha_{\rm FS}}{\tau \ln{\Lambda_c}}\sum_j \frac{n_j}{n_e}Z_{j}^2\left( 0.35 + 0.20\ln{p} \right) \equiv  \tau^{-1} p \left( \phi_{\text{br}0} + \phi_{\text{br}1}\ln{p}\right)
        \label{eq:bremsthralung}
    \end{equation}
    and $\alpha_{\rm FS}$ is the fine structure constant. The screening and radiation effectively increase the friction at large momenta, which will prevent runaway electrons from reaching arbitrarily large energies when given a long enough time to accelerate. 
    
    Equation \eqref{eq:full_kinetic_equation} describes the momentum space dynamics of the runaway phenomena, however it does not include any terms allowing for spatial transport. \cite{helander2000suppression} amended the kinetic description by adding the radial component of the diffusion operator in cylindrical geometry, characterised by the phase-space dependent diffusion coefficient $D$. Including this in our formulation we obtain the full kinetic equation of interest here,
    \begin{equation}
        \begin{aligned}
            \deri{f}{t} &= \frac{1}{p^2}\deri{}{p} \left[\left( -\xi \frac{E}{\tau} + p\nu_s + F_{\text{br}} + \frac{p\gamma}{\tau_{\text{syn}}}\left(1 - \xi^2\right)\right) p^2 f\right]\\
            &+ \deri{}{\xi}\left[ \left(1 - \xi^2\right)\left( -\frac{E/\tau}{p}f + \frac{1}{2}\nu_D\deri{f}{\xi} \right) - \frac{\xi\left(1 - \xi^2\right)}{\tau_{\text{syn}}\gamma} f \right] + \diff{D}{f}.
        \end{aligned}
        \label{eq:full_kinetic_equation_with_diffusion}
    \end{equation}
    In the next section we formulate a general expression for the change in the runaway avalanche growth rate resulting from such a finite $D$.

     \section{Reduced avalanche growth rate and effective critical electric field}
\label{sec:homogenous}
 The appearance of the diffusion term in the kinetic equation adds another dimension to the problem, a radial one, compared to the standard avalanche growth rate calculation \citep{jayakumar1993collisional,RosenbluthPutvinski1997}. We develop an approximate solution by taking advantage of a separation of time scales, following the approach outlined by \cite{helander2000suppression}. The avalanche generates secondary electrons with momentum predominantly close to the critical momentum for the runaway process $p_c$\footnote{The avalanche source term is $\frac{1}{p^2}\deri{}{p}\frac{1}{\gamma - 1} \sim \frac{1}{p^5}$ for low momenta, and therefore does not extend far into the runaway region.} and the electron is accelerated from this region up to relativistic momenta on the short timescale ${\tau_{\text{acc}} \sim 
 \tau/E}$. The transport timescale represented by $D$ will typically be longer than this acceleration time, so the diffusion will not be strong enough to significantly reduce or alter the generation process. However, the timescale of the avalanche growth is significantly longer, namely ${\gamma_r^{-1} \sim 2 \loglambda \;\tau_{\text{acc}}}$  \citep{jayakumar1993collisional} and thus the spatial diffusion may be expected to have a substantial impact on the avalanche. 

To this end, the momentum space is divided into a low energy region, $p < p_*$, where all the runaway generation occurs and the effects of radial diffusion are neglected, and a high energy region, $p > p_*$, where all the radial transport takes place. After this division in momentum space, the high energy region is modelled as source free and the generation of runaway electrons is modelled as a flux through the lower boundary in momentum space at $p_*$. Furthermore, the theory is reduced to only a single momentum-space coordinate in the high energy region.

This was done neglecting the effect of radiation in \citep{helander2000suppression}, by recognising that runaway electrons often have small pitch-angles, $\xi \approx 1$ and so expanding the collision operator assuming $\pperp \ll \ppar$, where $\ppar$ and $\pperp$ are the projections of the momentum along and perpendicular to the magnetic field line, respectively.
 The rate of change of the distribution function integrated over perpendicular momenta ${\cal F} = \int d^2\vec{p}_{\perp} f = \int_0^{\infty} d\pperp\; 2\pi\pperp f$  can then be obtained by   integrating the kinetic equation \eqref{eq:full_kinetic_equation_with_diffusion}, leading to
    \begin{equation}
        \dt{\cal{F}} + \frac{E - 1}{\tau} \derip{\cal{F}} = \diff{D(p)}{\cal{F}},
        \label{eq:non_radiative_diffe_equation}
    \end{equation}
    where the diffusion coefficient for particles travelling purely along the magnetic field line is used to first order. This is the kinetic description of the runaway electrons given in equation $(12)$ of~\citep{helander2000suppression}.
    The synchrotron radiation reaction force is however strongly dependent on the momentum perpendicular to the magnetic field line, as it is a consequence of the gyration around the field line, and a treatment of radiative effects needs to account for the pitch angle distribution of particles.
    
    Radiative effects are important close to the critical electric field where the acceleration from the electric field is close to being balanced by the radiation reaction forces, making the dynamics in the energy direction of momentum space comparatively slow. Therefore, as in~\cite{lehtinen1999monte,aleynikov2015theory,hesslow2018effect}, we consider the pitch-angle evolution to be a rapid process compared to the energy evolution dynamics and assume a steady-state in the pitch-angle distribution for a given $p$. This requires that the pitch-angle flux of particles vanishes, the condition following from the kinetic equation \eqref{eq:full_kinetic_equation_with_diffusion} as
    \begin{equation}
        0 = \left(1 - \xi^2\right)\left(-\frac{E/\tau}{p}f + \frac{1}{2}\nu_{D} \deri{f}{\xi}\right) - \frac{\xi(1 - \xi^2)}{\tau_{\text{syn}}\gamma}f.
        \label{eq:pitch_flux}
    \end{equation}
    Since $\tau_{\text{syn}} \gg \tau$ we formally neglect the effect of the synchrotron radiation on the pitch-angle distribution, retaining only the balance between the diffusive effect of pitch-angle scattering and the collimating effect of the electric field. 
    This can be used to solve for the pitch angle distribution and the distribution function may now be written as 
     \begin{equation}
        2\pi p^2 f(r, p, \xi, t) = \frac{A(p)}{2\sinh(A(p))}e^{A(p)\xi} F(r, p, t),
        \label{eq:pitch_distribution}
    \end{equation}
    where the reduced distribution function $F$ includes the $2\pi p^2$ of the momentum-space Jacobian, such that the radial density of runaway electrons is
    \begin{equation}
        \nRE(r, t) = \int_{p_*}^{\infty}dp\; F(r, p, t)
        \label{eq:def_nRE}
    \end{equation}
   and the inverse of $A(p) = 2E/\left(p\nu_D\tau\right)$ determines the extent of the distribution function in $\xi$. In the limit of large $p$, the pitch angle distribution is narrow in agreement with the treatment by \cite{helander2000suppression}, as $\nu_D \sim p^{-2}$ and therefore $A^{-1} \sim p^{-1}$. 
    
    Integrating the kinetic equation over pitch-angle, a reduced kinetic equation now accounting for radiation reaction forces and screening effects is obtained
    \begin{equation}
        \deri{F}{t} + \frac{1}{\tau}\derip{}\left(U(p) F\right) = \diff{\langle D \rangle_{\xi}}{F},
        \label{eq:radiative_diffe_equation}
    \end{equation}
    where the pitch averaged force $U(p)$ is 
    \begin{equation}
        U(p) = E\coth{A} - \tau \left[\frac{p\nu_{D}}{2} + p\nu_s  + F_{br} + \frac{p^2\gamma\nu_D}{\tau_{\text{syn}} E}\left(\coth{A} - \frac{1}{A}\right)\right],
        \label{eq:U_def}
    \end{equation}
    and the pitch-angle averaged diffusion coefficient is
    \begin{equation}
        \langle D \rangle_{\xi}(p) = \int_{-1}^{1}d\xi\; D(p, \xi)\frac{Ae^{A\xi}}{2\sinh{A}}.
        \label{eq:def_pitch_average_D}
    \end{equation}
    Note that for large $p$, neglecting screening effects, we recover the non-radiative result $U(p) \approx E\coth{A} - p \tau \nu_s \rightarrow E -1$.
    A qualitative difference between the models with and without radiative effects is the appearance of a momentum scale $\pmax$ where the pitch-angle averaged advection in momentum disappears, $U(\pmax) = 0$. This limits the energy of the relativistic particles and corresponds to the energy scale where radiation reaction forces balance the electric field acceleration. 
    
    As outlined by~\cite{helander2000suppression}, we impose the boundary condition that the particle flux through $p_*$ is given by the avalanche growth, $\gamma_r \nRE$, where $\gamma_r$ is the growth rate without the impact of diffusion. Given the structure of equation \eqref{eq:radiative_diffe_equation} this translates to a condition on $F$ as
    \begin{equation}
        F(p_*) = \frac{\tau \gamma_r}{U(p_*)}\nRE = \frac{\tau \gamma_r}{U(p_*)} \int_{p_*}^{\infty} dp\; F,
        \label{eq:F_boundary_condition}
    \end{equation}
    which is not a typical boundary condition as the value at the lower boundary in momentum space is dependent on the solution in the whole high energy region. A closed-form expression for the solution can be found for the simpler problem of radially-uniform plasma parameters in a quasi-steady state in terms of a Bessel mode expansion\footnote{The same type of solution is still possible with a radially dependent diffusion coefficient, as the radial part of the problem forms a Sturm–Liouville problem, however the expansion would no longer be in terms of Bessel functions but rather the eigenfunctions of the transport term in question.}. The effective strength of the diffusion experienced by each Bessel mode is scaled by the square of its inverse radial length scale $k_i = b_i / a$, where $b_i$ is the $i$:th root of the zeroth Bessel function $J_0(x)$ \citep{abramowitz1948handbook}. The solution is,

     \begin{equation}
        F(p, r, t) = \frac{1}{U(p)}\sum_{i = 1}^{\infty} c_i J_0(k_ir)\exp\left(\gamma_i t - \int_{p_*}^{p} dp' \frac{\tau}{U(p')}\left(\gamma_i + k_i^2 \langle D \rangle_{\xi}(p')\right)\right).
        \label{eq:distribution_function_radiation}
    \end{equation}
    The coefficients $c_i$ are determined by the initial condition, or seed profile of the avalanche process.
    The growth rate of the modes, $\gamma_i$, will be determined by the boundary condition \eqref{eq:F_boundary_condition}.
    
    \cite{helander2000suppression} considered only the first Bessel mode, $i = 1$. This gave a conservative estimate of the effect of diffusion as higher mode numbers have a smaller characteristic length scale so will experience a larger effect of diffusion, as noted above. Here we choose to retain all the modes, which would allow the runaway distribution function to be propagated in time. As a consequence of the orthogonality of the Bessel modes, the boundary condition can be projected on each mode separately, which decouples them from one another. The equation for $\gamma_i$ then follows from inserting \eqref{eq:distribution_function_radiation} in the boundary condition \eqref{eq:F_boundary_condition} as
    \begin{equation}
        1 = \int_{p_*}^{\pmax}dp\; \frac{\gamma_r \tau}{U(p)}\exp\left( -\int_{p_*}^{p}dp'\; \frac{\gamma_i \tau + \tau k_i^2\langle D \rangle_{\xi}}{U(p')} \right),
        \label{eq:radiative_int_equation}
    \end{equation}
    where the upper limit of the integration is $\pmax$ as no particles can gain energy larger than this.
    
    The theory for the pitch-angle distribution described above is valid for large $p$ -- which we are mostly concerned with here -- and not in general close to the critical momentum $p_c$, where $U(p_c) = 0$. If the free parameter $p_*$ is chosen close to $p_c$, the result will be sensitive to the choice. We can consistently minimise this impact of $p_*$ by expanding $U$ in large $p$, such that the theory still retains a $\pmax$, and safely set $p_* = p_c$ as a typical momentum scale for the onset of the runaway region.
    A large-$p$ expansion keeping terms of order $p^{-1}$ and larger gives 
    \begin{equation}
        \begin{aligned}
            U(p) &= E - \Bar{\nu}_{s0} + \frac{\tau\Bar{\nu}_{D0}^2}{2\tau_{\text{syn}}E^2} - \left(\Bar{\nu}_{s1} - \tau\frac{ \Bar{\nu}_{D0}\Bar{\nu}_{D1}}{\tau_{\text{syn}} E^2}\right)\ln{p} + \frac{\tau \Bar{\nu}_{D1}^{2}}{2\tau_{\text{syn}}E^2} \ln^2p\\
            &- \left(\phi_{br0} + \frac{\tau\Bar{\nu}_{D0}}{\tau_{\text{syn}}E}\right)p - \left(\phi_{br1} +  \frac{\tau\Bar{\nu}_{D1}}{\tau_{\text{syn}}E} \right)p\ln{p} - \left( \frac{1}{2} + \frac{\tau}{\tau_{\text{syn}} E}\right)\frac{\Bar{\nu}_{D0} + \Bar{\nu}_{D1}\ln{p}}{p}. 
        \end{aligned}
        \label{eq:def_U_large_p}
    \end{equation}
    
    Finally, a model for the avalanche growth rate without magnetic perturbations is needed. For continuity with our collision operator we will use the model by \cite{hesslow2019influence}, which incorporates the effect of screening by the evaluation of the collision frequencies at an effective critical momentum scale $p_c^{\rm eff}$, implicitly given by $p_c^{\rm eff} = \sqrt[4]{\Bar{\nu}_s(p_c^{\rm eff}) \Bar{\nu}_D(p_c^{\rm eff})} / \sqrt{E}$. Furthermore the model has an increased threshold field for the avalanche generation, $\Bar{E}_c^{\text{eff}}$, given by \cite{hesslow2018effect}. Here, the bar on  $\Bar{E}_c^{\text{eff}}$ indicates the critical electric field without the effect of radial diffusion, to distinguish from its value ${E}_c^{\text{eff}}$ when the radial diffusion is taken into account in section \ref{sec:Eceff}. The expression for the avalanche generation 
    is then 
    \begin{equation}
        \left(\deri{\nRE}{t} \right)^{\text{Aval}} \hspace{-0.1cm} = \gamma_r \nRE = \frac{n_e^{\text{tot}} / n_e}{\tau \ln{\Lambda} \sqrt{4 + \Bar{\nu}_s(p_c^{\rm eff}) \Bar{\nu}_D(p_c^{\rm eff})}} \left(E - \Bar{E}_{c}^{\text{eff}} \right) \nRE,
        \label{eq:standard_growth_rate}
    \end{equation}
    where $n_e^{\text{tot}}$ is the total number of electrons in the system ($\text{bound} + \text{free}$). Although we will use this model for the avalanche generation in the next section in order to consider its reduction due to radial transport, our method is agnostic to this choice and can be adapted to any avalanche description.

    \subsection{Reduced avalanche growth rate}
        As the radial transport allows for runaway electrons to be lost from the system, preventing these electrons from multiplying by the avalanche mechanism, the exponential growth rate of the Bessel modes, $\gamma_i$, will be reduced compared to the uncorrected value.
        The magnitude of the transport coefficients considered below reduces at large momentum, and in the avalanche distribution the particle density in phase space also decreases at large energies. Therefore, the problem is to a large extent determined by the dynamics at small energies. At these energies, where $p\ll \pmax$, the radiation reaction forces do not influence the problem significantly, and the acceleration dynamics is dominated by the electric field. For large field strength we have $U \approx E$. However, at the critical electric field for runaway generation, $\Bar{E}_c^{\text{eff}}$, the runaway generation and the advection in momentum space fall to zero. A linear interpolation between these regions gives $U \approx E - \Bar{E}_c^{\text{eff}}$.
        
\begin{figure}
            \centering
            \includegraphics[width=\textwidth]{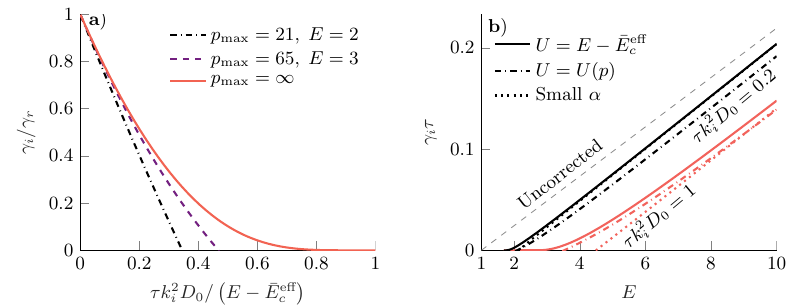}
            \caption{Reduction of the avalanche growth rate in a fully
              ionised plasma with $Z_{\text{eff}} = 1$, $T = 10$ eV,
              $n_e = 10^{20}\text{m}^{-3}$ and $B = 3$ T based on the
              numerical solution of equation
              \eqref{eq:radiative_int_equation} with the functional
              form of the diffusion coefficient from equation
              \eqref{eq:diff_coefficient}. \textbf{a}) The relative
              correction of the growth rate as a function of diffusion
              strength for different electric field strengths. The
              limit $\pmax = \infty$ corresponds to the theory with $U
              = E - \Bar{E}_c^{\text{eff}}$. \textbf{b}) The corrected
              growth rate as a function of electric field strength. At
              large electric field strength the offset of the
              corrected growth rate depends on the diffusion strength
              $\tau k_i^2 D_0$, which is expected to be around unity
              in an ITER-sized machine with normalized magnetic
              perturbation level of $\delta B/B\simeq 10^{-4}$.}
            \label{fig:correction_growth_rate}
        \end{figure}

        The uncorrected growth rate in equation \eqref{eq:standard_growth_rate} scales with electric field strength as $ E - \Bar{E}_{c}^{\text{eff}}$, which was just noted to be the approximate dependence of $U$ at low momentum  ($p\ll \pmax$), such that the prefactor of the exponent in \eqref{eq:radiative_int_equation} does not significantly vary with electric field strength at low $p$. The impact of the diffusion at low momentum is then characterised by 
        \begin{equation}
            \alpha = \tau k_i^2 \langle D \rangle_{\xi} \big/ (E - \Bar{E}_{c}^{\text{eff}})
        \end{equation}
        and the dependence of the corrected growth rate on this parameter is shown in figure \ref{fig:correction_growth_rate}a, which is obtained by solving equation \eqref{eq:radiative_int_equation} numerically. In the limit of small $\alpha$ the effect of diffusion is rather well parametrised by this normalised ratio of diffusion strength to the electric field acceleration. However, the correlation is lost when the growth rate is strongly reduced and approaches zero. Qualitatively, this can be understood as the effect of $\pmax$ and the high energy particles, 
        as a finite $\pmax$ limits the number of energetic particles that can contribute to the avalanche without being significantly affected by the diffusion. The effect is demonstrated in figure \ref{fig:correction_growth_rate}a by the case $U = E - \Bar{E}_c^{\text{eff}}$ where $\pmax$ is formally infinite and the growth rate remains slightly above zero for a relatively wide range of diffusion strengths\footnote{Formally, the growth rate cannot be corrected down to zero in the theory where $U = E - \Bar{E}_c^{\text{eff}}$ if the $D$ decays asymptotically in momentum space faster than $p^{-1}$. The latter is the marginal case where it is impossible if $\alpha_0 < 1$ where $\alpha \sim \alpha_0 / p$ asymptotically.}. This limit is approached generally at large electric field strength. 
        
       In figure \ref{fig:correction_growth_rate}a a diffusion coefficient of the following form was used,  
        \begin{equation}
            \langle D \rangle_{\xi}(p) = D_0 \frac{v}{c \gamma} = D_0 \frac{p}{1 + p^2}.
            \label{eq:diff_coefficient}
        \end{equation}
        The motivation for this expression is to capture the expected low energy behaviour where the diffusion is proportional to the particle velocity along the magnetic field line \citep{rechester1978electron}, as well as the expected effects of orbit decorrelation at high energy, as described for example in \cite{hauff2009runaway}. The latter authors made an estimate for $D_0$ in the small Kubo number limit, such that $D \simeq v_B^2 \tau_{\|}$, where the radial velocity of the particles $v_{B} = v_{\|} \delta B / B $ is due to the projection of the motion along the perturbed field line, with $\delta B$ the root mean square of the magnetic perturbation amplitude. The parallel correlation time is assumed to be set by the particle motion through the perturbed poloidal magnetic field structure, $\tau_{\|} = \lambda_{\|}/v_{\|} \simeq \pi q R / v_{\|}$, where $\lambda_{\|}$ is the parallel connection length and $q$ is the safety factor. Therefore, the estimate of $D_0$ is 
        \begin{equation}
            D_0 \simeq \pi q R \left(\delta B / B \right)^2 c .
        \end{equation}
         The strength of the diffusion is parametrised in this paper by $\tau k_i^2 D_0$ which is thus related to the magnetic perturbation level as follows, 
        \begin{equation}
            \tau k_i^2 D_0 \approx 3.14\; b_i^2 \frac{R[{\rm m}] q}{a[{\rm m}]^2 n_{e,20} \ln{\Lambda}} \left(10^{4} \delta B / B \right)^2,
            \label{eq:diffusion_strength}
        \end{equation}
        where $R[{\rm m}]$ and $a[{\rm m}]$ are the major and minor radii in meters, respectively, and $n_{e,20}$ is the electron density in units of $10^{20} \; \rm m^{-3}$. Consequently, for standard ITER parameters without any material injection, $R = 6.2$ m, $a = 2$ m, $n_e = 10^{20}\; \rm m^{-3}$, $\ln{\Lambda} \approx 15$ and $q \approx 1$, we have $\tau k_1^2 D_0 \approx  2 (\delta B / B)^2 \times 10^8$ for the least suppressed mode ($b_1 \approx 2.4$).

        In absolute units, the uncorrected growth rate scales linearly with the electric field strength and in figure \ref{fig:correction_growth_rate}b we see that the corrected growth rate also shows a linear relation with the electric field strength for large fields. The corrected growth rate is offset from the uncorrected one because the inverse of the characteristic diffusion parameter $\alpha$ and the uncorrected growth rate depend similarly on the electric field. 
        This can be seen by expanding equation \eqref{eq:radiative_int_equation} in small $\alpha$ yielding\footnote{This formula can be arrived at by using integration by parts and change the momentum variable $q = p_* + \int_{p_*}^{p}dp'\; \left(E - \Bar{E}_{c}^{\text{eff}}\right)/U(p')$ in the intermediate steps. The transformation maps the problem to a theory without radiative effects.}
        
        \begin{equation}
            \gamma_i = \gamma_r - \int_{p_*}^{\pmax}dp\; \gamma_r \frac{\tau k_i^2 \langle D \rangle_{\xi}}{U(p)}\exp\left(-\int_{p_*}^{p}dp'\; \frac{\gamma_r \tau }{U(p')} \right),
        \end{equation}
        where the second term is almost independent of $E$ if $\pmax$ is large, since then $U \approx E-{E}_c^{\text{eff}}$ for the values of $p$ with the largest contribution to the integral. This approximation is compared to the full numerical solution in figure \ref{fig:correction_growth_rate}b where the offset from the uncorrected result is evident for large $E$ and small $\alpha$.
        
    \subsection{Effective critical electric field}\label{sec:Eceff}
        As the radial transport reduces the growth rate, the critical electric field strength for net generation of runaway electrons may increase. In the current decay phase of the disruption the electric field stays close to the critical electric field and the current decay rate is proportional to its value \citep{BreizmanAleynikov2017Review,hesslow2018effect}. Therefore the critical electric field has direct relevance for disruption mitigation strategies. 
        
        The mode least suppressed by the transport is the lowest order mode, with index $i=1$, and therefore can be expected to dominate the runaway profile in the late stages of the disruption. We therefore choose to define the effective critical electric field $E_c^{\text{eff}}$ as the field strength at which the growth rate of the first mode vanishes, namely $\gamma_1 = 0 $. 
        
\begin{figure}
            \centering 
            \includegraphics[width=\textwidth]{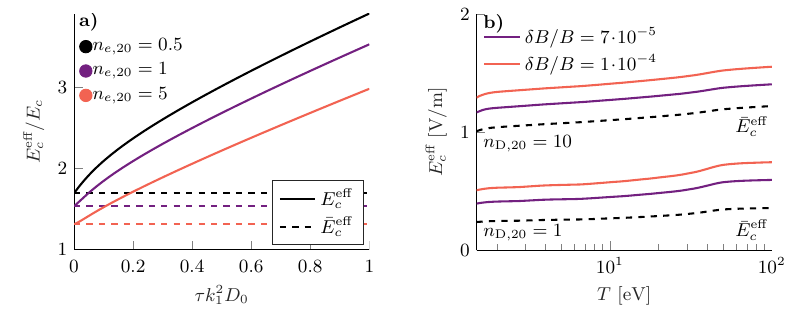}
            \caption{\textbf{a}) Critical electric field as a function of diffusion strength calculated using the momentum space dependent diffusion coefficient given in equation \eqref{eq:diff_coefficient}.  The critical electric field for a net avalanche gain in the presence of spatial transport (solid lines) is enhanced compared to the theory without transport (dashed lines). A linear relation between effective critical field strength and diffusion strength is found for large diffusion strengths. The plasma is fully ionised with effective charge $Z_{\text{eff}} = 1$, a temperature of $T = 10$ eV and magnetic field strength $B = 3$ T. The large $p$-expansion for $U$ has been used. \textbf{b}) Critical electric field as a function of temperature, in a plasma with deuterium density $n_{\text{D}} = 10^{21}\;\text{m}^{-3}$ (upper three curves) or $n_{\text{D}} = 10^{20}\;\text{m}^{-3}$ (lower three curves), and neon density $n_{\text{Ne}} = 10^{19}\;\text{m}^{-3}$ in both cases, where the respective ionisation states for all temperatures $T$ are determined assuming equilibrium based on the ADAS coefficients of ionisation and recombination. The strength of the diffusion is characterised by equation \eqref{eq:diffusion_strength} with minor and major radii $a = 2$\,m and $R = 6.2$ m respectively, with a safety factor $q$ of order unity. 
            }
            \label{fig:critical_electric_field}
        \end{figure}

        Figure \ref{fig:critical_electric_field}a shows numerical solutions for the critical electric field, obtained from equation \eqref{eq:radiative_int_equation} under the constraint $\gamma_1 = 0$ in fully ionised plasmas with different densities. For large diffusion strengths, when the effective critical electric fields are relatively large, a linear relation between diffusion strength and electric field is found.  This follows naturally for large electric fields $E$ in \eqref{eq:radiative_int_equation}, $U \approx E - \Bar{E}_{c}^{\text{eff}}$ and $\pmax$ is at large enough energy scales not to be relevant, as the condition $\gamma_1 = 0$ translates to a condition on $\alpha$. Given this linear relation in diffusion coefficient, the critical electric field strength is expected to be quadratic in the magnetic perturbation level, $\delta B / B$. 
        
        Figure \ref{fig:critical_electric_field}b shows the critical effective field as a function of temperature, which introduces screening effects at the lowest temperatures, where some electrons remain bound to ions. The ionisation states are determined here by assuming equilibrium based on the ADAS coefficients of ionisation and recombination\footnote{http://www.adas.ac.uk}. 
        
        In absolute units, the correction to the effective critical field strength increases with temperature, primarily due to the increase in free electrons which raises $E_c$. Massive material injection will also raise the density and so the critical electric field for runaway generation. We see from figure \ref{fig:critical_electric_field}b that this effect can be combined with the effect of spatial diffusion to further raise $E_c^{\text{eff}}$. However unlike massive material injection, where changes to $E_c^{\text{eff}}$ are linked to changes in the electron density, the correction based on magnetic perturbations is only weakly dependent on the density (as long as perturbation strength is treated independent of plasma density), in absolute units. This weak dependence follows as $\tau \gamma_r / E$ is density independent and therefore any sensitivity originates from only $\Bar{E}_c^{\text{eff}}$ and the large-$p$ dependence of $U$. Further, as was shown by \cite{hesslow2018effect},  the effect of partial screening on the critical electric field $\Bar{E}_c^{\text{eff}}$ was to raise it to the order of $E_c^{\text{tot}}$, which has the same form as the usual Connor-Hastie expression $E_c$, but with the combined density of free and bound electrons instead of only the density of free electrons. Therefore, only a weak dependence of temperature is seen in figure \ref{fig:critical_electric_field}b, as the total number of electrons are kept fixed in these simulations. The two mechanisms for increasing the effective electric field (screening and magnetic perturbations) can therefore be combined. We note, however, that $\Bar{E}_c^{\text{eff}}$ is only weakly dependent on the temperature and ionisation state.
        
        The theory discussed so far, in which radial variations in the plasma have been neglected, allows for self-consistent analytic solutions to the distribution function and an understanding of the dependencies of the growth rate correction due to transport. However, in a tokamak disruption, the electric field dynamics is essential and will vary spatially through its dependence on plasma properties. Therefore in the next section we take a perturbative approach to solving equation \eqref{eq:radiative_diffe_equation}, to include effects due to radial plasma variation.  \section{Transport in an inhomogeneous plasma}
\label{sec:inhomogeneous}
    During the current quench of a disruption the flux surfaces are not completely stochastic. Instead, they often exhibit a mixed magnetic topology consisting of intact flux-surfaces, magnetic islands and stochastic regions. In such circumstances, the transport will no longer be well described by the expression given by \cite{rechester1978electron} and a more general transport model consisting of spatially dependent diffusive and advective components is often formulated, based on particle following simulations \citep{Papp_2015,Sarkimaki_2016}. Assuming cylindrical symmetry, the transport term on the right hand side of equation \eqref{eq:radiative_diffe_equation} becomes 
   \begin{equation}
        \frac{1}{r} \deri{}{r} r \left(- \langle V \rangle_{\xi} + \langle D \rangle_{\xi} \deri{}{r} \right) F, 
    \end{equation}
    where $\langle V \rangle_{\xi}$ is the pitch-angle average of the
    radial component of the advection coefficient defined equivalently
    to \eqref{eq:def_pitch_average_D}.  Fundamentally the above
    transport term conserves particle number, which is a property not
    guaranteed by an approximate perturbative solution. Therefore,
    this conservation property will be imposed on the solution to
    prevent anomalous losses of particles. Note, that the conservation of
    particle number is local, and particles can  be lost at the
    edge.
  
    The approach to solving the kinetic equation in the high energy region with a momentum space dependent diffusion coefficient in the previous section, by means of a Bessel mode expansion, breaks down when a radial dependence is introduced in the plasma parameters.
    To include the effects of radially varying plasma parameters we instead perform an expansion in small radial transport compared to the electric field acceleration $\alpha \ll 1$. The full form of $U$ gives $U=0$ in the vicinity of $\pmax$, so the transport would not be subdominant for such momenta. We therefore treat this with a simplified approach, taking $U = E - \Bar{E}_c^{\text{eff}}$ which is similar to the advection in momentum space given by \cite{helander2000suppression},  modified to include the effects of an increased critical electric field due to screening. In this model, the advection in momentum-space does not vanish, and in the previous section, we saw that the full $U$  approaches this in the limit of large electric field. As this model does not treat the effects of radiation correctly, it may be seen as a better approximation significantly below $\pmax$. As the assumed transport coefficients and particle density decrease with momentum in avalanche dominated scenarios, these lower energy scales are anyway expected to dominate the transport here. We will also demonstrate explicitly the weak sensitivity of the results to $p_*$.
    
    Given this model for $U$, the zeroth order solution given by neglecting the transport, so the solutions at different radii are independent,  is
    \begin{equation}
        F_0(p, r, t) = \nRE(r, t) \frac{\gamma_r \tau }{E - \Bar{E}_{c}^{\text{eff}}} e^{- \gamma_r \tau \left(p - p_*\right) / \left(E - \Bar{E}_{c}^{\text{eff}}\right)},
        \label{eq:F0}
    \end{equation}
    which is consistent with the incoming runaway electron flux $\gamma_r \nRE$ and respects the definition of $\nRE$ in \eqref{eq:def_nRE}. Under the assumption that the radial transport is small this distribution may be used to evaluate the transport term. Integrating equation \eqref{eq:radiative_diffe_equation} over $p$ then gives 
    \begin{equation}
        \deri{\nRE}{t} + \frac{1}{r}\deri{}{r}\left( r \Gamma_0 \right) = \gamma_r \nRE,
        \label{eq:conservative_RE_equation}
    \end{equation}
    where the term on the right represents the source of runaway electrons and 
    \begin{equation}
        \Gamma_0 = \int_{p_*}^{\infty}\; \left(\langle V \rangle_{\xi} F_0 - \langle D \rangle_{\xi} \deri{}{r}F_0\right)\; dp
    \end{equation}
    is the radial flux of the runaway electrons. Using equation \eqref{eq:F0} for $F_0$ to evaluate $\Gamma_0$ results in a form $\Gamma_0 = \Bar{\Gamma}_0 \nRE + \Tilde{\Gamma}_0 \partial_r \nRE$ and equation  \eqref{eq:conservative_RE_equation} can be stably solved numerically using a method based on the Crank–Nicholson scheme. This transport model can be incorporated into any suitable runaway simulation to similarly capture the effects of transport due to magnetic perturbations. In the following subsections the transport model will be integrated with the electric field evolution, but it should be noted that the momentum space shape of the distribution function, equation \eqref{eq:F0}, is insensitive to the electric field strength, due to the appearance of the ratio $\gamma_r / (E - \Bar{E}_c^{\text{eff}})$, which is consistent with the neglect of the temporal evolution of $E$ in its derivation.

    \subsection{Simplified disruption simulations including the effect of radial transport}\label{eq:simplified_disruptions}
   Under normal operation of a tokamak there are radial gradients in the temperature and plasma current, both of which contribute to a radially varying electric field as the plasma is suddenly cooled in a disruption. The subsequent evolution of the electric field is described by the induction equation, which for a plasma with only radial variations is \citep{smith2006runaway}
    \begin{equation}
 \frac{1}{r}\deri{}{r}r\deri{E_{\|}}{r} = \mu_0 \deri{j_{\|}}{t},
        \label{eq:inductive_equation}
    \end{equation}
    where 
   $E_{\|}$ and $j_{\|}$ are the electric field strength and the current density along the plasma cylinder.

    The time evolution of the electric field is dependent on its radial profile, as the current $j_{\|}$ has both an Ohmic and a runaway component: ${j_{\|} = \sigma_{\text{sp}} E_{\|} + j_{\text{RE}}}\; {\approx \sigma_{\text{sp}} E_{\|} + ec\nRE}$ where $\sigma_{\text{sp}}$ is the Spitzer conductivity \citep{spitzer1953transport}.
    Accordingly, the radial profile of runaway generation also plays a crucial part in understanding the electric field evolution. Furthermore, the avalanche growth rate is proportional to the electric field strength, for fields large compared to the critical one, such that the cumulative generation is highly dependent on the evolution of the electric field. Consequently, for a self-consistent treatment of both the electric field dynamics and runaway generation, it is of the utmost importance to be able to treat the runaway generation in a region of space with an electric field gradient. Such computations can be carried out within the {\sc go}-framework \citep{smith2006runaway,Feher-PPCF-11,Vallhagen-JPP-20} or similar codes. 
    Introducing the effect of transport losses into such a model, including the effects of impurities and allowing for partial screening, would allow us to quantify the reduction given by transport of the total number of runaway electrons at the end of a disruption. 
    
    To achieve this, we have extended the {\sc go}-framework to solve the coupled equations \eqref{eq:conservative_RE_equation} and \eqref{eq:inductive_equation}. The runaway generation and transport are described by equation \eqref{eq:conservative_RE_equation}, while the runaway electron dynamics couples to the electric field evolution through the current term in equation \eqref{eq:inductive_equation}. The right hand side of the former includes primary sources of runaway electrons: Dreicer generation, tritium decay and Compton sources.   When avalanching dominates, as is the case for high current devices, the flux $\Gamma_0$ derived using only the  avalanche source should be valid as it describes the momentum space distribution of the majority of the population. Finite-aspect-ratio effects on the generation will be neglected here, as recent work by \cite{McDevitt_2019} indicate that their effect is negligible at the high densities and electric fields that we will consider here.
    
    In the {\sc go}-framework, the Dreicer generation is evaluated using a neural network \citep{hesslow2019evaluation}, trained on kinetic simulations with  {\sc code} \citep{landreman2014numerical}, using the collision operator that includes the effect of partially ionized impurities given by \cite{HesslowJPP}.  $\beta$-decay of tritium will also result in a source of runaway electrons, which in the deuterium-tritium phase of operation is expected to be the dominant source of seed electrons in ITER in the absence of hot-tail electrons \citep{martin2017formation}. Furthermore, neutrons produced in the fusion reactions will activate the wall which in turn will emit $\gamma$-photons. Through Compton scattering events between the $\gamma$-photons and the bulk electrons, runaway electrons can be generated \citep{martin2017formation,Vallhagen-JPP-20}. In the simulations presented here, we neglect the hot-tail generation occurring in a rapidly cooling plasma. This generation occurs during the thermal quench which is typically associated with large magnetic fluctuations and corresponding transport. By neglecting the hot-tail seed we implicitly assume that the transport during the thermal quench is large enough to lead to the prompt loss of these runaways. 
        
  The {\sc go}-framework has the capability to compute the plasma temperature evolution from the energy balance between heat diffusion, Ohmic heating, line radiation, bremsstrahlung losses and ionisation, as presented by \cite{Vallhagen-JPP-20}. However, in the initial phase of the disruption, the energy loss is expected to be dominated by the MHD contribution, due to its strong temperature scaling $\sim T^{5/2}$ \citep{ward1992impurity}. This phase is, for simplicity, modelled as an exponential drop in temperature until the temperature of the inner part of the plasma drops to $\sim 100$ eV, with the form  
        \begin{equation}
            T(r, t) = T_{\text{f}}(r) + \big(T_{\text{i}}(r) - T_{\text{f}}(r)\big) e^{- t / t_0},
            \label{eq:T_evolution}
        \end{equation}
        where $t_0$ is the time constant for the thermal quench and $T_{\text{i}}$ and $T_{\text{f}}$ are the initial and final temperatures, respectively. This mode of the temperature evolution uses a flat final temperature profile $T_{\text{f}} = 50$ eV and is used for $6$ ms with a time constant of $t_0 = 1$ ms. After this time the temperature is evolved based on the energy balance. The ionisation states in the background plasma are evolved in time based on the ADAS coefficients for ionisation and recombination.

    \subsection{Simulations of ITER-like disruptions with uniform perturbations}
        To investigate the large scale effect of radial transport, in tokamak disruption scenarios where the runaway generation is expected to be dominated by the avalanche mechanism, an ITER-like case with deuterium and neon injection was simulated using the {\sc go}-framework. The parameters considered are the same as those used by \cite{martin2017formation} and \cite{Vallhagen-JPP-20}: initial plasma current $I_p(t = 0) = 15$ MA,  minor and major radii $a = 2$ m and $R = 6.2$ m, respectively, initial electron,  deuterium and tritium densities $n_{e0} = 2 n_{D0} = 2 n_{T0} = 10^{20}\; \text{m}^{-3}$. The simulation was initiated with one dimensional profiles in temperature $T_{\text{e}} = 20\big[1 - \left(r/a\right)^2\big]$ keV and current density $j_{\|}(t = 0) = j_0 \big[1 - \left( r / a \right)^2\big]^{0.41}$, where $j_0$ is chosen so that the current integrates to 15~MA.
        
\begin{figure}
            \centering
            \includegraphics[width=\textwidth]{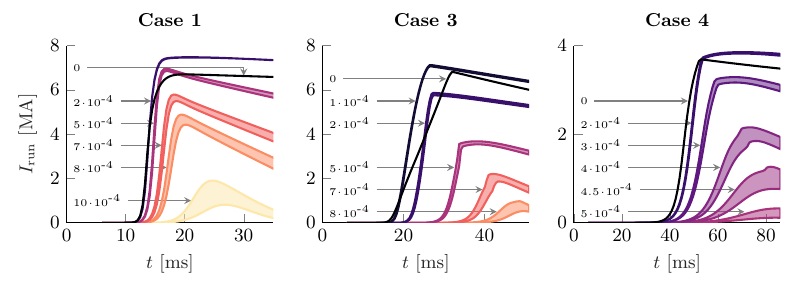}
            \caption{Evolution of current carried by runaway electrons in an ITER-like disruption in the presence of magnetic perturbations. The magnitude of $\delta B / B$ is shown by the text and colour in the figure. Three cases of injected material are considered: A pure neon injection with density $n_{\text{Ne}} / n_{e0} = 1$ \mbox{(\textbf{Case 1})}, and two cases with the same amount of injected neon $n_{\text{Ne}} / n_{e0} = 0.08$, but different amount of injected deuterium $n_{\text{D}} / n_{e0} = 40$ \mbox{(\textbf{Case 3})} and $n_{\text{D}} / n_{e0} = 7$ \mbox{(\textbf{Case 4})}, respectively.
            The coloured area corresponds to $p_*$ in the range $0.1 - 1$. }
            \label{fig:cases}
        \end{figure}

       At the start of the simulations we assume a rapid injection of deuterium and neon, with respective densities $n_{\text{D}}$ and $n_{\text{Ne}}$, where the impurity is distributed uniformly throughout the plasma in the neutral state. 
       The current evolution for three such scenarios with different combinations of injected neon and deuterium is demonstrated in figure \ref{fig:cases}, for a set of radially constant magnetic perturbation levels and a momentum space dependent diffusion coefficient of the form \eqref{eq:diff_coefficient}. The three cases considered are the same as those investigated in \cite{Vallhagen-JPP-20} and denoted both here and there as Case 1, Case 3 and Case 4\footnote{Case 2 described in \citep{Vallhagen-JPP-20} does not result in a complete thermal collapse.}. In each of these cases the injected material is large enough to induce a complete thermal quench: Case 1 ($n_{\rm Ne}=1\times 10^{20}\,\rm m^{-3}$, $n_{\rm D}=0$), Case 3 (${n_{\rm Ne}=8\times 10^{18}\,\rm m^{-3}}$, $n_{\rm D}=4\times 10^{21}\,\rm m^{-3}$) and Case 4 ($n_{\rm Ne}=8\times 10^{18}\,\rm m^{-3}$, $n_{\rm D}=7\times 10^{20}\,\rm m^{-3}$). In the absence of perturbations large runaway currents were obtained in all of these three cases, even without hot-tail generation. 
       
       Interestingly,  the maximum runaway current increases for small transport coefficients ($\delta B / B \simeq 2\cdot 10^{-4}$) compared to the baseline case of no radial transport, as the runaway electron seed is radially flattened by the transport. This agrees with previous results by \cite{Feher-PPCF-11}. However, for large enough perturbations we note a reduction of the maximal current carried by the runaway electrons. How large the reduction is depends on the particular scenario. Case 4, with a combination of moderate neon and deuterium injection shows the largest reduction.
       
       The effectiveness of the radial transport in modifying the runaway evolution is closely related to the time scale of the current evolution. Generally, the longer the time scale of the current quench the more pronounced is the effect of transport, as particles have more time to be transported out of the plasma. Due to this effect Case 4 has a slower growth rate of runaway electrons than in Cases 1 and 3, diffusion therefore having a larger impact, as can be noted in figure \ref{fig:cases}.
       
\begin{figure}
                \centering 
                \includegraphics[width=6.3cm]{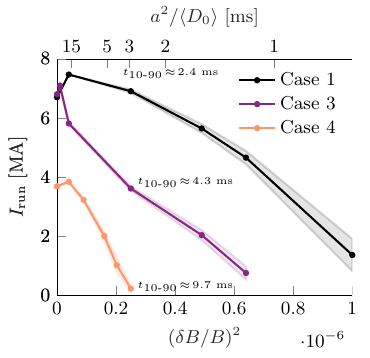}
            \caption{Runaway current in ITER-like disruptions in the presence of magnetic perturbations. The maximum current carried by the runaways is shown against the square of the magnetic perturbation level - proportional to the transport coefficient. The upper axis label shows the diffusion time scale $a^2/\langle D_0\rangle$. The shaded area corresponds to the range of $p_*$ shown in figure \ref{fig:cases}. The time for the runaway current to rise from  $10\%$ to $90\%$ of its maximum value, $t_{10\text{-}90}$, is shown in the figure for $\delta B / B = 2\cdot 10^{-4}$.
            }
            \label{fig:reduction_lines}
        \end{figure}

       Figure \ref{fig:reduction_lines} shows the maximum runaway current, just before the onset of the dissipation phase where the plasma current carried by the runaway electrons decays, in the three cases as a function of $(\delta B/B)^2$. We note that almost full suppression of the runaway current can be achieved in Case 4, for a normalised perturbation $\delta B/B \simeq 5\cdot 10^{-4}$. We also show the time it takes for the runaway current to rise from $10\%$ to $90\%$ of its maximum value, denoted by $t_{10\text{-}90}$, for $\delta B / B = 2\cdot 10^{-4}$. Clearly, Case 4 has considerably longer $t_{10\text{-}90}$ than the other two and this is the main reason for the larger suppression.  The diffusion time scale can be estimated to be $t_{\rm diff}=a^2/\langle D_0\rangle \simeq a^2/ (\pi \langle q \rangle R c) \left(\delta B / B \right)^{-2} $, and is $17\;\rm ms$ for $\langle q \rangle \simeq 1$ and $\delta B / B = 2\cdot 10^{-4}$. Here, $\langle\cdots\rangle$ denotes a volume average. We note that the uncertainty in $p_*$ influences the result, but the effect of the magnetic perturbation is clearly dominant.
    
        \begin{figure}
            \centering 
            \includegraphics[width=6.45cm]{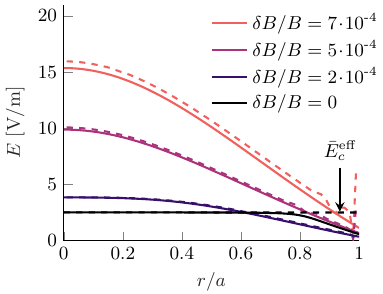}
            \caption{Electric field after $50$ ms in Case 3 shown in figure \ref{fig:cases} (solid line), together with the approximate runaway plateau electric field obtained from setting the local growth of the runaway density to zero in equation \eqref{eq:conservative_RE_equation} (dashed line) using the radial profile of runaway electrons from the simulation.}
            \label{fig:currents2}
        \end{figure}
        
        When the runaway plateau phase is reached in the final stages of the disruption the loss of plasma current is dominated by the loss of runaway electrons. If the time derivative of the Ohmic current is neglected in equation \eqref{eq:inductive_equation}, combining with equation \eqref{eq:conservative_RE_equation} yields an approximate equation for the electric field in the runaway plateau, 
        \begin{equation}
            \frac{1}{r}\deri{}{r}r\deri{E_{\|}}{r} \approx \mu_0 c e \left( \gamma_r(E_{\|}) \nRE - \frac{1}{r}\deri{}{r}\left(r \Gamma_0\right) \right).
            \label{eq:E_plateau}
        \end{equation}
        Based on this expression the decay of the plasma current is 
        \begin{equation}
            \deri{I_p}{t} = 
            \frac{2\pi a}{\mu_0}
            \deri{E_{\|}}{r}\Big|_{r = a} \approx 2\pi
            e c \left(\int_{0}^{a}dr\; r \gamma_{r}(E_{\|}) \nRE  - a \Gamma_0(r = a)\right),
            \label{eq:current_decay}
        \end{equation}
        for a given radial profile of the runaway electron density. This approximation recovers the electric field structure in the current decay phase to a large extent, however an even simpler consideration can be made where the local growth of runaway electrons is set to zero, in equation \eqref{eq:conservative_RE_equation}, which in terms of equation \eqref{eq:E_plateau} corresponds to neglecting the left hand side of the equation. Using the expression for the uncorrected growth rate under consideration here, equation \eqref{eq:standard_growth_rate}, yields an explicit expression for the electric field given a profile $\nRE$.  Figure \ref{fig:currents2} shows the electric field obtained in the simulations corresponding to the cases shown in figure \ref{fig:cases}, together with the electric field strength which zeros the local growth of runaway electrons. The electric field profiles are seen to agree with one another in regions with small gradients in $E$ and especially in the central part of the plasma, which is related to the neglected term from equation \eqref{eq:E_plateau}.  
        Therefore, the electric field is generally higher than the threshold field for runaway generation $\Bar{E}_c^{\text{eff}}$ in the centre of the plasma, where there is a balance between the generation and transport. However, close to the edge the current density is low enough so that the prefactor on the source term in equation \eqref{eq:E_plateau} is small enough to allow a significant deviation below $\Bar{E}_c^{\text{eff}}$. Furthermore, in the cases with large transport, the electric field in the central region is not as flat as $\Bar{E}_c^{\text{eff}}$. Instead it has a similar functional form to the runaway current profile, indicating that it is the transport which dominates the electric field in the plateau phase. In these situations, the electric field is highly dependent on the profile of runaway electrons, which in turn depends on the full temporal evolution of the system, and in particular the transport coefficients. This suggests that approaches where the current decay phase is described by $\Bar{E}_c^{\text{eff}}$ are only valid if the transport is negligible, otherwise the coupled dynamics with the runaway electrons must be considered. 
        
     \section{Runaway dynamics in the presence of artificial resonant  perturbations} \label{sec:iter}
\begin{figure}
                \centering                 
                \includegraphics[width=0.5\textwidth]{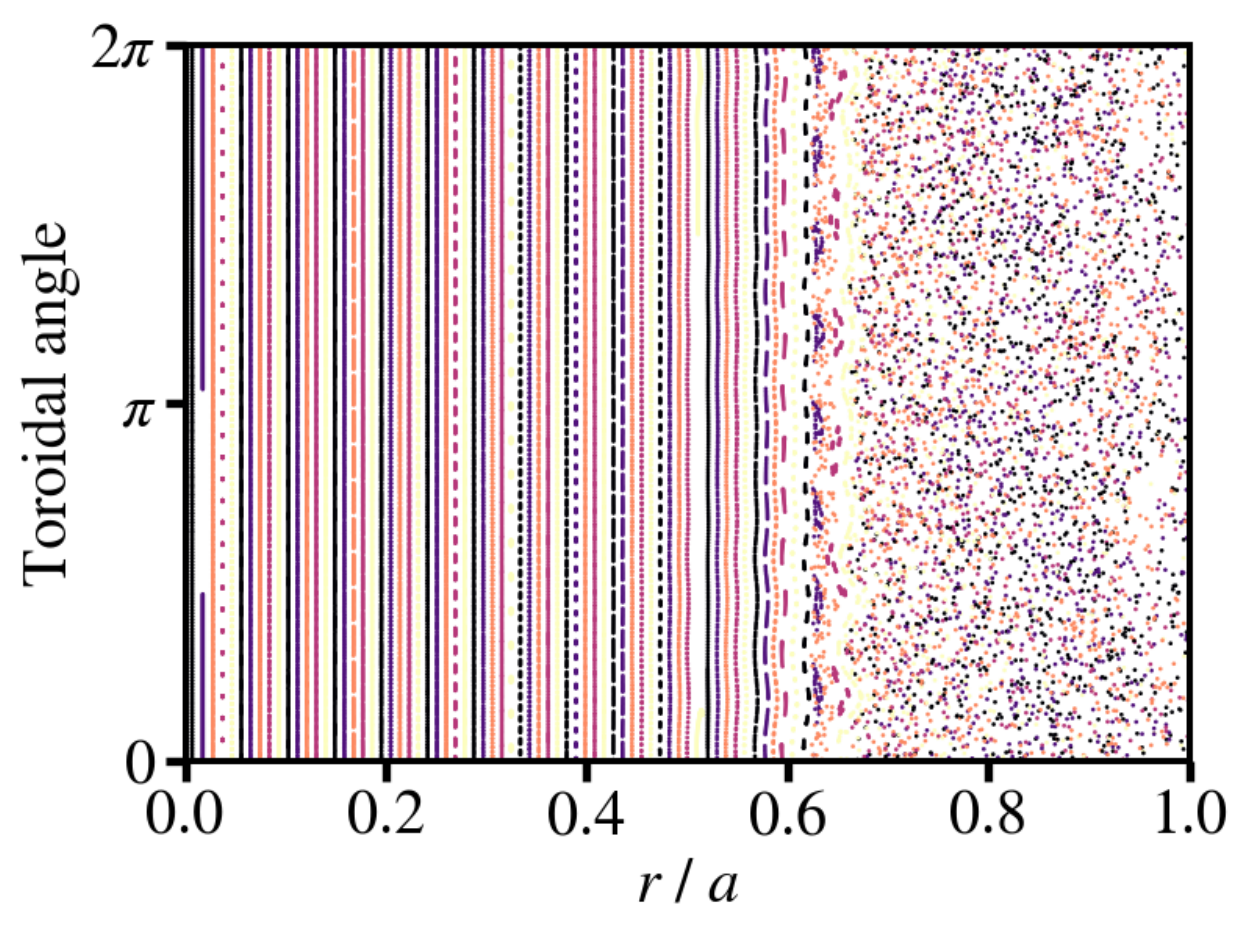}
               \caption{Magnetic field Poincar\'e-plot at the outer mid-plane for an ITER current flat-top equilibrium perturbed with artificial resonant magnetic perturbations according to \eqref{eq:pert}. The stochastic region begins at $r/a\approx 0.6$ ($q=1$ surface is at $r/a\approx 0.5$).}
    \label{fig:poincare}
    \end{figure}

    The magnetic field is expected to become fully stochastic at the end of the thermal quench, after which it begins to heal during the current quench. To mimic the conditions during the current quench in ITER, we choose the 15~MA / 5.3~T baseline scenario \citep{Parail_2013}
    %\footnote{https://www.iter.org/technical-reports}
    and introduce artificial resonant magnetic perturbations at the plasma edge to create a stochastic layer. We have chosen the pre-disruption current flat top equilibrium for this exercise as obtaining realistic current quench equilibrium would require dedicated MHD modelling. The introduced perturbations are stationary and have a helical structure,
    \begin{equation}
    \delta \mathbf{B} = \nabla\times\sum_{n,m}\alpha_{nm}(\rho)\cos(n\zeta- m \theta - \phi_{nm})\mathbf{B},
    \label{eq:pert}
    \end{equation}
    where $(\rho,\;\theta,\;\zeta)$ are the radial, poloidal, and toroidal Boozer coordinates, respectively, $\mathbf{B}$ is the unperturbed field and the phase $\phi_{nm}$ is chosen to be random. This method is the same as used in \cite{Sarkimaki_2020}, and also here the total perturbation consists of several modes with low mode numbers $(n,m \lesssim 20)$. The mode eigenfunctions are Gaussians,
   	\begin{equation}
   	\alpha_{nm} = \exp\left(\frac{(r-r_{nm})^2}{2\sigma^2}\right),
	\end{equation}    
that peak at the corresponding resonance $r_{nm}$ and all have the same width $\sigma=0.03$~m which is large enough for the modes to overlap and create a continuous stochastic region. The perturbation level is set to $\delta B/B \approx 10^{-3}$ at which significant runaway transport is expected \citep{helander2000suppression}. The resulting field is illustrated in figure~\ref{fig:poincare}.
    The transport coefficients are evaluated numerically with the orbit-following code ASCOT5 \citep{ascot5}. Markers representing guiding centers of collisionless electrons were traced in the perturbed field, and their radial position was recorded for each orbit. As particles with finite orbit width oscillate radially during their orbit, the radial position was always recorded at the same poloidal position (at the outer mid-plane) so that all changes in the radial position were due to the transport alone. No collisions, electric field or radiation reaction force was present in the simulation to isolate the transport due to the magnetic field perturbations and to keep momentum constant in order to calculate momentum dependent coefficients. 
    
    The transport coefficients are evaluated from the recorded radial positions as
    \begin{align}
        \label{eq:drift evaluation}
        V &= \frac{1}{N}\sum_{i = 1}^N\left<\frac{\Delta r}{\Delta t}\right>_i,\\
        \label{eq:diffusion evaluation}
        D &= \frac{1}{N}\sum_{i = 1}^N\left<\frac{(\Delta r)^2}{\Delta t}\right>_i,
    \end{align}
    where the brackets denote an average over all collected data points for a marker $i$, $\Delta t$ is the orbit circulation time, $\Delta r$ is the change in radial position between subsequent recordings, and the sum is taken over all $N$ markers that were traced. This scheme is similar to the one originally presented by \cite{boozer1981monte}. At the edge markers are lost within a few orbits, making estimates~\eqref{eq:drift evaluation} and~\eqref{eq:diffusion evaluation} unreliable, and so the coefficients are evaluated from the loss-time distribution using the method described in \citep{Sarkimaki_2016,Sarkimaki_2020}. This latter method is used if more than half of the markers are lost.
    
    \begin{figure}
        \centering 
          \includegraphics[width=\textwidth]{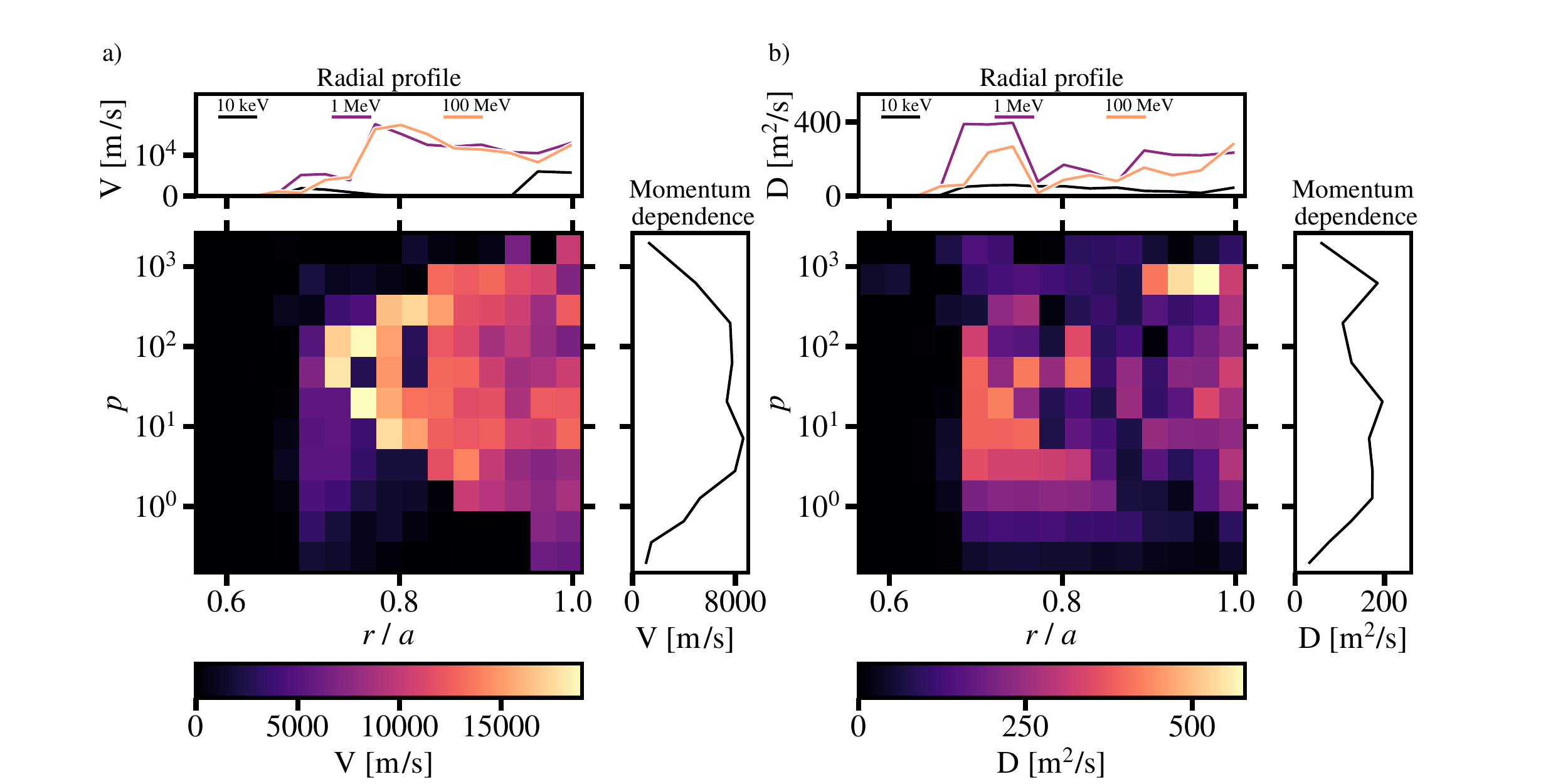}
        \caption{Numerically evaluated advection (a) and diffusion (b) coefficients for the transport due to the stochastic field corresponding to the case shown in Fig.~\ref{fig:poincare}. The 2D plots show the radial and momentum dependence of the advection and diffusion coefficients for a fixed pitch $p_\parallel/p= 0.99$. Radial profiles at different energies are shown at the top. At the side, general momentum dependence is illustrated with a mean value calculated over each radial position.}
            \label{fig:coefficients}
    \end{figure}
    
    Markers are simulated for $2\times10^{-5}$~s which corresponds to approximately 100 orbit transit times. The simulation time has to be long enough as early on the particle orbits are correlated and the motion is not diffusive. However, longer simulation time decreases the radial resolution of the evaluated coefficients as $\Delta r \approx \sqrt{2Dt}$. The markers have identical radial position, pitch and momentum but a random toroidal location. The transport coefficients to be used in the {\sc go} simulation are found by repeating the orbit-following simulation with different values of radius, pitch, and momentum. In these simulations, the phase space is divided into 15 radial, 11 momentum, and 10 pitch slots (covering the passing particle regime). For each volume element 200 markers are simulated to calculate the coefficients at that point.
    The resulting advection and diffusion coefficients are shown in figure~\ref{fig:coefficients} for a fixed pitch, as the coefficients show no strong pitch dependence as long as the particles are passing. Radially the transport is almost uniform in the region where the field is stochastic (recall figure~\ref{fig:poincare}) while the momentum dependence shows decreasing transport for higher energies due to the finite orbit width effects \citep{hauff2009runaway,Sarkimaki_2020}.
    
    In the inner region of the plasma ($r / a < 0.58$), the runaway electrons are not transported as the flux surfaces are intact, and markers initiated in the stochastic region will not be transported into this region. To properly capture this effect in a simulation with the {\sc go}-framework, a reflective boundary condition was imposed at the first complete flux surface, and no transport could occur between the regions. However, the regions are still coupled through the electric field evolution.

In scenarios where the runaway generation is dominant in the central part (such as in Case 1 and Case 4) the stochastic plasma edge is not expected to affect the runaway dynamics significantly. By using the advection and diffusion coefficients shown in figure~\ref{fig:coefficients}, and simulating an ITER-like scenario with material injection corresponding to Case 4, we find that the maximum runaway current is reduced only marginally, from 3.7 MA to 3.5 MA.
    To illustrate a case when the effect of a stochastic plasma edge is more pronounced, we consider Case 3, where in the absence of radial losses an off-axis final runaway profile is found. Therefore the transport should have a larger impact in this case compared to scenarios with an on-axis final current profile, where a larger part of the runaway electrons are generated in the non-transporting region. 
 
\begin{figure}
        \centering 
        \includegraphics[width=7cm]{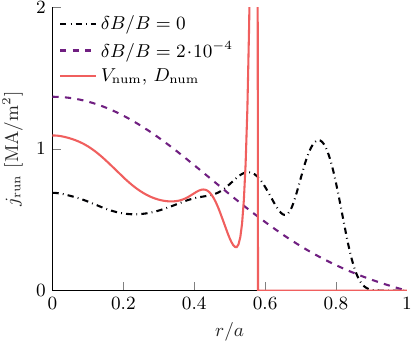}
        \caption{Radial profiles of the runaway current after $45$ ms in the ITER-like disruption simulation of Case 3 without magnetic perturbations (dash-dotted), with radially constant magnetic perturbations $\delta B/B=2\cdot 10^{-4}$ (dashed) and with the coefficients presented in figure \ref{fig:coefficients} (solid). In the latter case, a strong current sheet develops at the interface to the stochastic region. }
        \label{fig:profiles_ITER}
    \end{figure}

    Figure \ref{fig:profiles_ITER} shows the radial profiles of the runaway current after $45$ ms in the ITER-like disruption simulation of Case 3. The maximum runaway current, just before the dissipation phase, in the absence of radial transport due to magnetic perturbations is $7$~MA, with a constant $\delta B / B$ is $5.8$ MA and with the coefficients presented in figure \ref{fig:coefficients} is $4.6$ MA. Without magnetic perturbations, the profile of runaway electron density has an off-axis maximum. This is due to strong radiative losses, leading to significant plasma cooling, and corresponding efficient runaway generation in the outer part of the plasma, as was pointed out by \cite{Vallhagen-JPP-20}. Note, that such off-axis current profiles may become MHD unstable. The resulting magnetic activity may act to mitigate the build up of runaways.

In the presence of magnetic perturbations electrons can diffuse and
this results in a final runaway current profile that is peaked
on-axis.  However, in the case with the stochastic edge, with
transport coefficients shown in figure~\ref{fig:coefficients}, the
transport in the edge region is strong enough to prevent any
significant build up of runaway electrons there. The increase
of the transport at the edge results in only partial reduction of the
total runaway current, as the runaway population is free to build up
in the centre of the plasma. In the confined inner region strong
gradients in the current density can form, which is evident in the
simulation. At the transition from the confined region to the
stochastic one (at $r/a=0.58$) a discontinuity is formed in the
current profile (however not in the electric field), as particles in
the stochastic outer region are continuously transported away, but in
the confined region they are free to build up, eventually forming a
current sheet.
 
\begin{figure}
        \centering
        \includegraphics[width=\textwidth]{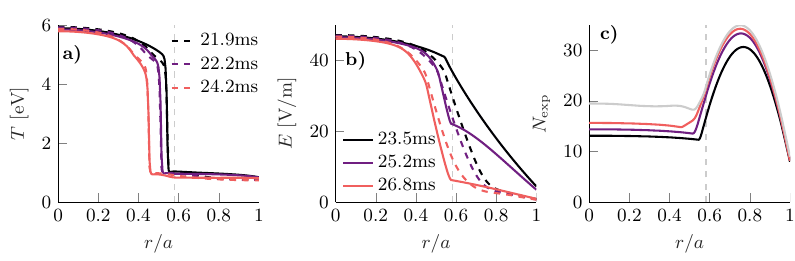}
  
        \caption{Radial profiles from the ITER-like disruption simulation in Case 3, with the transport coefficient presented in figure \ref{fig:coefficients} (solid lines) and without transport of runaway electrons (dashed lines), at subsequent time slices. Radial profiles of \textbf{a)} temperature, \textbf{b)} electric field and \textbf{c)} the number of e-foldings defined in equation \eqref{eq:e_foldings}.
        The time slices were chosen to highlight the formation of the current sheet in the case with transport, and are identified in panel b). The dashed lines were taken at times such that the positions of the cold front were matched. The extra (gray) line in c) gives the number of e-foldings at the start of the current decay phase. The vertical dashed line shows the onset of the stochastic region.}
        \label{fig:case3evol}
    \end{figure}

    The radial profiles of the temperature, electric field and number of e-foldings (the time-integral of the runaway growth rate) are shown in figure \ref{fig:case3evol} for Case 3,  at a few time slices. The vertical dashed line denotes the radial position for the transition between confined and stochastic regions.  Figure \ref{fig:case3evol}a shows that, both with and without perturbations, the plasma is divided into two regions by a cold front, with an inner region with a temperature of about 6 eV, and an outer region with a temperature as low as about 1 eV. At such low temperatures a large fraction of the deuterium recombines, and this leads to an increased avalanche multiplication of the seed runaway electrons in the outer region, see figure \ref{fig:case3evol}c, quantified by the number of e-foldings\footnote{$\exp \left(N_{\text{exp}}\right)$ is the factor by which the avalanche mechanics amplifies the seed in a non-transporting model.}, 
    \begin{equation}
        N_{\text{exp}}(r) = \int_{0}^{t}dt'\; \gamma(t', r).
        \label{eq:e_foldings}
    \end{equation}
    The avalanche production continues throughout the simulation, but is counteracted by the strong radial transport in this region. 
    
    The formation and strength of the current sheet seen in figure \ref{fig:profiles_ITER} is a result of the interaction between runaway transport and strong diffusion of the electric field.
    The location is tightly connected to that of the cold front, which propagates in from the plasma edge in the later stages of the simulation.
    In the scenario without perturbations, the cold front propagates inwards faster. Figure \ref{fig:case3evol} compares the evolution of the electric field in the two scenarios after the cold front has crossed out of the stochastic region, at times when the cold front has reached the same position, to highlight the dynamics behind the current sheet formation. In the case with transport there is less conversion from Ohmic to runaway current in the outer parts of the plasma, so a significantly larger electric field is maintained in the outer region, as is seen in figure \ref{fig:case3evol}b. When the conversion starts in the inner regions, a sharp change in the gradient of the electric field develops at the interface to the stochastic region, which enhances the diffusion of the strong electric field    in the inner region. This results in a larger avalanche multiplication of the seed runaway electrons    in the boundary between the regions,
    which is demonstrated in figure \ref{fig:case3evol}c.
Despite the strong amplification in the outer region, the transport prevents a significant number of runaway electrons building up, and so the current sheet is formed. 

Another way of thinking of this phenomenon is to consider the effect of the runaway electrons on the electric field evolution. As the transport in the outer region is strong enough to prevent a runaway build up, the runaways do not affect the electric field evolution - this is akin to the assumption of a trace electron population used in the estimate by \cite{RosenbluthPutvinski1997} - which leads to a very large amplification factor. However, the coupled dynamics must be considered in the presence of a large runaway population, and the amplification saturates as the current carried by the runaways in the inner region here approaches the Ohmic current. Then the current sheet is formed at the interface between the two regions. Such a current profile is likely to be very unstable, and this could affect the magnetic equilibrium and lead to magnetic perturbations penetrating deeper into the plasma core, giving further runaway mitigation. Such a study is beyond the scope of the current paper. \section{Conclusions}
During tokamak disruptions the magnetic field lines can be severely distorted from their usual confining structure. The magnetic topology evolves in time, being almost fully stochastic during the thermal quench, and often displaying a mixed topology of intact flux-surfaces, magnetic islands and stochastic regions during the current quench. As runaway electrons travel rapidly along the tokamak magnetic field lines, their evolution during a disruption can be strongly affected by such magnetic perturbations. This introduces the possibility for radial losses of runaway electrons to offset the avalanche growth of the population, preventing the formation of a high current, potentially damaging, runaway electron beam.

In this paper, we have presented a model which generalises previous treatments of the effects of radial transport due to the interaction of runaway electrons with magnetic field perturbations on the runaway evolution. We continue to take advantage of the separation of timescales in the runaway generation dynamics, between the acceleration to relativistic energies after a knock-on collision and the characteristic avalanche population growth time. This allows us to neglect the effect of transport due to magnetic perturbations on the generation process and focus on solving the kinetic equation in the high energy limit, which simplifies the collision operator. The extension here allows for a generalised pitch-angle distribution formed by rapid pitch-angle scattering at high energy, the impact of radiation reaction and the presence of partially ionised impurity atoms. The effect of including radiation is to introduce an upper limit in momentum space, so particles are prevented from reaching very high energies where they would be well confined.

In particular, we have determined an expression which can be used to correct the growth rate of the runaway electron population by the avalanche  mechanism. This takes the form of the solution of an integral equation.
The introduction of radial transport raises the effective critical electric field for avalanche generation because even though particles can be kicked into the runaway region through the avalanche mechanism, they can be lost due to spatial diffusion resulting in no net gain of runaways. The increase in $\Eceff$ is weakly dependent on the plasma density, unlike the case of massive material injection.

The derivation of the integral equation for the growth rate corrections is only valid when radial variations in the plasma
are neglected. To treat non-homogeneous plasmas, a perturbation approach in small radial transport has been used to estimate the radial flux of runaway electrons. By computing the radial fluxes instead of an effective (local) growth rate the formulation is particle conserving. This flux can be included in general runaway simulation frameworks. The formulation was used here in a simplified disruption simulation for ITER-like plasmas, where an induction equation was used to give the self-consistent time evolution of the electric field in the presence of the runaways.

We find that in scenarios with a moderate amount of impurity and deuterium injection, the runaway current can be suppressed for perturbation levels of the order of ${\delta B/B\sim 5\cdot 10^{-4}}$, which is about an order of magnitude higher than the perturbation level measured in fixed magnetic field experiments where the current was scanned  \citep{Gill_2002}. Furthermore, it is difficult to fully dissipate the runaway electrons without significant transport in the centre where most of the runaway electrons are generated.  Earlier results investigating the potential of employing edge-localized mode (ELM) coils for runaway suppression show that perturbations created at the plasma edge are generally not sufficient for runaway suppression in ITER \citep{Papp_2011PPCF}. Perturbations at the edge might have an effect on scenarios with an off-axis runaway current profile, which can arise in the case of massive material injection \citep{Vallhagen-JPP-20}. We investigated such a scenario, using orbit-following simulations with ASCOT5 \citep{Sarkimaki_2020} to determine the diffusion coefficients for energetic electrons in an ITER-like plasma with a stochastic region at the plasma edge.  Disruption simulations with these diffusion coefficients show that the final runaway current can be reduced, but not suppressed completely, in agreement with the conclusions of earlier work. 

The analysis presented in this paper is valid when the distribution function is in quasi-steady state, i.e.~the runaway generation is balanced by transport, and the transport can be described by an advection-diffusion model. This assumption is not valid during the thermal quench phase in the disruption or in the presence of major magnetic islands in which a fraction of runaways could remain confined. 
Furthermore, a kinetic effect lacking in the model is the impact of diffusion on the avalanche generation dynamics at momentum scales close to the critical one. We anticipate that this effect could be small compared to the effects investigated so far, as the runaway electrons spend a comparatively short amount of time close to the critical momentum. Analytical progress in this direction would require the addition of a radial dimension in the full kinetic calculation with a source term to treat the dynamics close to the critical momentum, then the development of a solution to the kinetic equation valid for large momenta in the same calculation.  Numerical progress, on the other hand, could be made directly by implementing the radial transport in kinetic frameworks. This would have the added benefit of capturing the pitch-angle dynamics in the presence of pitch-angle dependent transport coefficients, an effect which has not been considered in the present work.

The authors are grateful to I.~Pusztai, G.~Papp, L.~Hesslow, M.~Hoppe and A.~Tinguely for fruitful discussions.  This work was supported by the European Research Council (ERC) under the European Union’s Horizon 2020 research and innovation programme (ERC-2014-CoG grant 647121) and the Swedish Research Council (Dnr.~2018-03911).

\bibliographystyle{jpp}

\bibliography{ref}

\begin{thebibliography}{58}
\expandafter\ifx\csname natexlab\endcsname\relax\def\natexlab#1{#1}\fi
\def\au#1{#1} \def\ed#1{#1} \def\yr#1{#1}\def\at#1{#1}\def\jt#1{\textit{#1}}
  \def\bt#1{#1}\def\bvol#1{\textbf{#1}} \def\vol#1{#1} \def\pg#1{#1}
  \def\publ#1{#1}\def\arxiv#1{#1}\def\org#1{#1}\def\st#1{\textit{#1}}

\bibitem[Abramowitz \& Stegun(1948)]{abramowitz1948handbook}
{\sc \au{Abramowitz, M.} \& \au{Stegun, I.~A.}} \yr{1948} {\em Handbook of
  mathematical functions with formulas, graphs, and mathematical tables\/}, ,
  \vol{vol.~55}.  \publ{US Government printing office}.

\bibitem[Aleynikov \& Breizman(2015)]{aleynikov2015theory}
{\sc \au{Aleynikov, P.} \& \au{Breizman, B.~N.}} \yr{2015}  \at{Theory of two
  threshold fields for relativistic runaway electrons}.  \jt{Physical Review
  Letters}  \bvol{114}~(15),  \pg{155001}.

\bibitem[Bandaru {\em et~al.\/}(2019)Bandaru, Hoelzl, Artola, Papp \&
  Huijsmans]{Bandaru2019}
{\sc \au{Bandaru, V.}, \au{Hoelzl, M.}, \au{Artola, F.~J.}, \au{Papp, G.} \&
  \au{Huijsmans, G. T.~A.}} \yr{2019}  \at{Simulating the nonlinear interaction
  of relativistic electrons and tokamak plasma instabilities: Implementation
  and validation of a fluid model}.  \jt{Phys. Rev. E}  \bvol{99},
  \pg{063317}.

\bibitem[Boozer \& Kuo-Petravic(1981)]{boozer1981monte}
{\sc \au{Boozer, A.~H.} \& \au{Kuo-Petravic, G.}} \yr{1981}  \at{Monte {C}arlo
  evaluation of transport coefficients}.  \jt{The Physics of Fluids}
  \bvol{24}~(5),  \pg{851--859}.

\bibitem[Breizman {\em et~al.\/}(2019)Breizman, Aleynikov, Hollmann \&
  Lehnen]{Breizman_2019}
{\sc \au{Breizman, B.~N.}, \au{Aleynikov, P.}, \au{Hollmann, E.~M.} \&
  \au{Lehnen, M.}} \yr{2019}  \at{Physics of runaway electrons in tokamaks}.
  \jt{Nuclear Fusion}  \bvol{59}~(8),  \pg{083001}.

\bibitem[Breizman \& Aleynikov(2017)]{BreizmanAleynikov2017Review}
{\sc \au{Breizman, B.~N.} \& \au{Aleynikov, P.~B.}} \yr{2017}  \at{Kinetics of
  relativistic runaway electrons}.  \jt{Nuclear Fusion}  \bvol{57}~(12),
  \pg{125002}.

\bibitem[Connor \& Hastie(1975)]{connor1975relativistic}
{\sc \au{Connor, J.~W.} \& \au{Hastie, R.~J.}} \yr{1975}  \at{Relativistic
  limitations on runaway electrons}.  \jt{Nuclear fusion}  \bvol{15}~(3),
  \pg{415}.

\bibitem[Embr{\'e}us {\em et~al.\/}(2016)Embr{\'e}us, Stahl \&
  F{\"u}l{\"o}p]{embreus2016effect}
{\sc \au{Embr{\'e}us, O.}, \au{Stahl, A.} \& \au{F{\"u}l{\"o}p, T.}} \yr{2016}
  \at{Effect of bremsstrahlung radiation emission on fast electrons in
  plasmas}.  \jt{New Journal of Physics}  \bvol{18}~(9),  \pg{093023}.

\bibitem[Feh\'{e}r {\em et~al.\/}(2011)Feh\'{e}r, Smith, F\"{u}l\"{o}p \&
  G\'{a}l]{Feher-PPCF-11}
{\sc \au{Feh\'{e}r, T.}, \au{Smith, H.~M.}, \au{F\"{u}l\"{o}p, T.} \&
  \au{G\'{a}l, K}} \yr{2011}  \at{Simulation of runaway electron generation
  during plasma shutdown by impurity injection in {ITER}}.  \jt{Plasma Physics
  and Controlled Fusion}  \bvol{53},  \pg{035014}.

\bibitem[F{\"u}l{\"o}p {\em et~al.\/}(2020)F{\"u}l{\"o}p, Helander, Vallhagen,
  Embreus, Hesslow, Svensson, Creely, Howard \&
  Rodriguez-Fernandez]{fulop2020effect}
{\sc \au{F{\"u}l{\"o}p, T.}, \au{Helander, P.}, \au{Vallhagen, O.},
  \au{Embreus, O.}, \au{Hesslow, L.}, \au{Svensson, P.}, \au{Creely, A.~J.},
  \au{Howard, N.~T.} \& \au{Rodriguez-Fernandez, P.}} \yr{2020}  \at{Effect of
  plasma elongation on current dynamics during tokamak disruptions}.
  \jt{Journal of Plasma Physics}  \bvol{86}~(1).

\bibitem[Gill {\em et~al.\/}(2002)Gill, Alper, de~Baar, Hender, Johnson,
  Riccardo \& contributors to~the EFDA-JET~Workprogramme]{Gill_2002}
{\sc \au{Gill, R.~D.}, \au{Alper, B.}, \au{de~Baar, M.}, \au{Hender, T.~C.},
  \au{Johnson, M.~F.}, \au{Riccardo, V.} \& \au{contributors to~the
  EFDA-JET~Workprogramme}} \yr{2002}  \at{Behaviour of disruption generated
  runaways in {JET}}.  \jt{Nuclear Fusion}  \bvol{42}~(8),  \pg{1039--1044}.

\bibitem[Granetz {\em et~al.\/}(2014)Granetz, Esposito, Kim, Koslowski, Lehnen,
  Mart\'{\i}n-Sol\'{\i}s, Paz-Soldan, Rhee, Wesley, Zeng \& Group]{Granetz}
{\sc \au{Granetz, R.~S.}, \au{Esposito, B.}, \au{Kim, J.~H.}, \au{Koslowski,
  R.}, \au{Lehnen, M.}, \au{Mart\'{\i}n-Sol\'{\i}s, J.~R.}, \au{Paz-Soldan,
  C.}, \au{Rhee, T.}, \au{Wesley, J.~C.}, \au{Zeng, L.} \& \au{Group,
  ITPA~MHD}} \yr{2014}  \at{An {ITPA} joint experiment to study runaway
  electron generation and suppression}.  \jt{Physics of Plasmas}
  \bvol{21}~(7),  \pg{072506}.

\bibitem[Hauff \& Jenko(2009)]{hauff2009runaway}
{\sc \au{Hauff, T.} \& \au{Jenko, F.}} \yr{2009}  \at{Runaway electron
  transport via tokamak microturbulence}.  \jt{Physics of Plasmas}
  \bvol{16}~(10),  \pg{102308}.

\bibitem[Helander {\em et~al.\/}(2000)Helander, Eriksson \&
  Andersson]{helander2000suppression}
{\sc \au{Helander, P.}, \au{Eriksson, L-G.} \& \au{Andersson, F.}} \yr{2000}
  \at{Suppression of runaway electron avalanches by radial diffusion}.
  \jt{Physics of Plasmas}  \bvol{7}~(10),  \pg{4106--4111}.

\bibitem[Helander \& Sigmar(2005)]{helander2005collisional}
{\sc \au{Helander, P.} \& \au{Sigmar, D.~J.}} \yr{2005} {\em Collisional
  transport in magnetized plasmas\/}.  \publ{Cambridge University Press}.

\bibitem[Hesslow {\em et~al.\/}(2018{\natexlab{{\em a\/}}})Hesslow, Embr\'eus,
  Hoppe, DuBois, Papp, Rahm \& F\"ul\"op]{HesslowJPP}
{\sc \au{Hesslow, L.}, \au{Embr\'eus, O.}, \au{Hoppe, M.}, \au{DuBois, T.~C.},
  \au{Papp, G.}, \au{Rahm, M.} \& \au{F\"ul\"op, T.}} \yr{2018{\natexlab{{\em
  a\/}}}}  \at{Generalized collision operator for fast electrons interacting
  with partially ionized impurities}.  \jt{Journal of Plasma Physics}
  \bvol{84}~(6),  \pg{905840605}.

\bibitem[Hesslow {\em et~al.\/}(2017)Hesslow, Embr{\'e}us, Stahl, DuBois, Papp,
  Newton \& F{\"u}l{\"o}p]{hesslow2017effect}
{\sc \au{Hesslow, L.}, \au{Embr{\'e}us, O.}, \au{Stahl, A.}, \au{DuBois,
  T.~C.}, \au{Papp, G.}, \au{Newton, S.} \& \au{F{\"u}l{\"o}p, T.}} \yr{2017}
  \at{Effect of partially screened nuclei on fast-electron dynamics}.
  \jt{Physical Review Letters}  \bvol{118}~(25),  \pg{255001}.

\bibitem[Hesslow {\em et~al.\/}(2019{\natexlab{{\em a\/}}})Hesslow,
  Embr{\'e}us, Vallhagen \& F{\"u}l{\"o}p]{hesslow2019influence}
{\sc \au{Hesslow, L.}, \au{Embr{\'e}us, O.}, \au{Vallhagen, O.} \&
  \au{F{\"u}l{\"o}p, T.}} \yr{2019{\natexlab{{\em a\/}}}}  \at{Influence of
  massive material injection on avalanche runaway generation during tokamak
  disruptions}.  \jt{Nuclear Fusion}  \bvol{59}~(8),  \pg{084004}.

\bibitem[Hesslow {\em et~al.\/}(2018{\natexlab{{\em b\/}}})Hesslow,
  Embr{\'e}us, Wilkie, Papp \& F{\"u}l{\"o}p]{hesslow2018effect}
{\sc \au{Hesslow, L.}, \au{Embr{\'e}us, O.}, \au{Wilkie, G.~J.}, \au{Papp, G.}
  \& \au{F{\"u}l{\"o}p, T.}} \yr{2018{\natexlab{{\em b\/}}}}  \at{Effect of
  partially ionized impurities and radiation on the effective critical electric
  field for runaway generation}.  \jt{Plasma Physics and Controlled Fusion}
  \bvol{60}~(7),  \pg{074010}.

\bibitem[Hesslow {\em et~al.\/}(2019{\natexlab{{\em b\/}}})Hesslow, Unnerfelt,
  Vallhagen, Embreus, Hoppe, Papp \& F{\"u}l{\"o}p]{hesslow2019evaluation}
{\sc \au{Hesslow, L.}, \au{Unnerfelt, L.}, \au{Vallhagen, O.}, \au{Embreus,
  O.}, \au{Hoppe, M.}, \au{Papp, G.} \& \au{F{\"u}l{\"o}p, T.}}
  \yr{2019{\natexlab{{\em b\/}}}}  \at{Evaluation of the {Dreicer} runaway
  generation rate in the presence of high-{$Z$} impurities using a neural
  network}.  \jt{Journal of Plasma Physics}  \bvol{85}~(6).

\bibitem[Hirvijoki {\em et~al.\/}(2015{\natexlab{{\em a\/}}})Hirvijoki, Decker,
  Brizard \& Embr{\'e}us]{hirvijoki2015guiding}
{\sc \au{Hirvijoki, E.}, \au{Decker, J.}, \au{Brizard, A.~J.} \&
  \au{Embr{\'e}us, O.}} \yr{2015{\natexlab{{\em a\/}}}}  \at{Guiding-centre
  transformation of the radiation--reaction force in a non-uniform magnetic
  field}.  \jt{Journal of Plasma Physics}  \bvol{81}~(5),  \pg{475810504}.

\bibitem[Hirvijoki {\em et~al.\/}(2015{\natexlab{{\em b\/}}})Hirvijoki,
  Pusztai, Decker, Embr{\'e}us, Stahl \& F{\"u}l{\"o}p]{hirvijoki2015radiation}
{\sc \au{Hirvijoki, E.}, \au{Pusztai, I.}, \au{Decker, J.}, \au{Embr{\'e}us,
  O.}, \au{Stahl, A.} \& \au{F{\"u}l{\"o}p, T.}} \yr{2015{\natexlab{{\em
  b\/}}}}  \at{Radiation reaction induced non-monotonic features in runaway
  electron distributions}.  \jt{Journal of Plasma Physics}  \bvol{81}~(5),
  \pg{475810502}.

\bibitem[Hollmann {\em et~al.\/}(2015)Hollmann, Aleynikov, F\"ul\"op,
  Humphreys, Izzo, Lehnen, Lukash, Papp, Pautasso, Saint-Laurent \&
  Snipes]{HollmannDMS}
{\sc \au{Hollmann, E.~M.}, \au{Aleynikov, P.~B.}, \au{F\"ul\"op, T.},
  \au{Humphreys, D.~A.}, \au{Izzo, V.~A.}, \au{Lehnen, M.}, \au{Lukash, V.~E.},
  \au{Papp, G.}, \au{Pautasso, G.}, \au{Saint-Laurent, F.} \& \au{Snipes,
  J.~A.}} \yr{2015}  \at{Status of research toward the {ITER} disruption
  mitigation system}.  \jt{Physics of Plasmas}  \bvol{22}~(2),  \pg{021802}.

\bibitem[Hollmann {\em et~al.\/}(2013)Hollmann, Austin, Boedo, Brooks, Commaux,
  Eidietis, Humphreys, Izzo, James, Jernigan, Loarte, Martin-Solis, Moyer,
  Munoz-Burgos, Parks, Rudakov, Strait, Tsui, Zeeland, Wesley \&
  Yu]{Hollmann2013}
{\sc \au{Hollmann, E.~M.}, \au{Austin, M.~E.}, \au{Boedo, J.~A.}, \au{Brooks,
  N.~H.}, \au{Commaux, N.}, \au{Eidietis, N.~W.}, \au{Humphreys, D.~A.},
  \au{Izzo, V.~A.}, \au{James, A.~N.}, \au{Jernigan, T.~C.}, \au{Loarte, A.},
  \au{Martin-Solis, J.}, \au{Moyer, R.~A.}, \au{Munoz-Burgos, J.~M.},
  \au{Parks, P.~B.}, \au{Rudakov, D.~L.}, \au{Strait, E.~J.}, \au{Tsui, C.},
  \au{Zeeland, M. A.~Van}, \au{Wesley, J.~C.} \& \au{Yu, J.~H.}} \yr{2013}
  \at{Control and dissipation of runaway electron beams created during rapid
  shutdown experiments in {DIII-D}}.  \jt{Nuclear Fusion}  \bvol{53},
  \pg{083004}.

\bibitem[Jayakumar {\em et~al.\/}(1993)Jayakumar, Fleischmann \&
  Zweben]{jayakumar1993collisional}
{\sc \au{Jayakumar, R.}, \au{Fleischmann, H.~H.} \& \au{Zweben, S.~J.}}
  \yr{1993}  \at{Collisional avalanche exponentiation of runaway electrons in
  electrified plasmas}.  \jt{Physics Letters A}  \bvol{172}~(6),
  \pg{447--451}.

\bibitem[Landreman {\em et~al.\/}(2014)Landreman, Stahl \&
  F{\"u}l{\"o}p]{landreman2014numerical}
{\sc \au{Landreman, M.}, \au{Stahl, A.} \& \au{F{\"u}l{\"o}p, T.}} \yr{2014}
  \at{Numerical calculation of the runaway electron distribution function and
  associated synchrotron emission}.  \jt{Computer Physics Communications}
  \bvol{185}~(3),  \pg{847--855}.

\bibitem[Lehnen {\em et~al.\/}(2009)Lehnen, Abdullaev, Arnoux, Bozhenkov,
  Jakubowski, Jaspers, Plyusnin, Riccardo \& Samm]{Lehnen2009}
{\sc \au{Lehnen, M.}, \au{Abdullaev, S.S.}, \au{Arnoux, G.}, \au{Bozhenkov,
  S.A.}, \au{Jakubowski, M.W.}, \au{Jaspers, R.}, \au{Plyusnin, V.V.},
  \au{Riccardo, V.} \& \au{Samm, U.}} \yr{2009}  \at{Runaway generation during
  disruptions in {JET} and {TEXTOR}}.  \jt{Journal of Nuclear Materials}
  \bvol{390-391},  \pg{740 -- 746}.

\bibitem[Lehnen {\em et~al.\/}(2015)Lehnen, Aleynikova, Aleynikov, Campbell,
  Drewelow, Eidietis, Gasparyan, Granetz, Gribov, Hartmann, Hollmann, Izzo,
  Jachmich, Kim, Ko{\v{c}}an, Koslowski, Kovalenko, Kruezi, Loarte, Maruyama,
  Matthews, Parks, Pautasso, Pitts, Reux, Riccardo, Roccella, Snipes, Thornton
  \& de~Vries]{Lehnen2015}
{\sc \au{Lehnen, M.}, \au{Aleynikova, K.}, \au{Aleynikov, P.B.}, \au{Campbell,
  D.J.}, \au{Drewelow, P.}, \au{Eidietis, N.W.}, \au{Gasparyan, Yu.},
  \au{Granetz, R.S.}, \au{Gribov, Y.}, \au{Hartmann, N.}, \au{Hollmann, E.M.},
  \au{Izzo, V.A.}, \au{Jachmich, S.}, \au{Kim, S.-H.}, \au{Ko{\v{c}}an, M.},
  \au{Koslowski, H.R.}, \au{Kovalenko, D.}, \au{Kruezi, U.}, \au{Loarte, A.},
  \au{Maruyama, S.}, \au{Matthews, G.F.}, \au{Parks, P.B.}, \au{Pautasso, G.},
  \au{Pitts, R.A.}, \au{Reux, C.}, \au{Riccardo, V.}, \au{Roccella, R.},
  \au{Snipes, J.A.}, \au{Thornton, A.J.} \& \au{de~Vries, P.C.}} \yr{2015}
  \at{Disruptions in {ITER} and strategies for their control and mitigation}.
  \jt{Journal of Nuclear Materials}  \bvol{463},  \pg{39 -- 48}.

\bibitem[Lehnen {\em et~al.\/}(2008)Lehnen, Bozhenkov, Abdullaev \&
  Jakubowski]{Lehnen_2008}
{\sc \au{Lehnen, M.}, \au{Bozhenkov, S.~A.}, \au{Abdullaev, S.~S.} \&
  \au{Jakubowski, M.~W.}} \yr{2008}  \at{Suppression of runaway electrons by
  resonant magnetic perturbations in {TEXTOR} disruptions}.  \jt{Phys. Rev.
  Lett.}  \bvol{100},  \pg{255003}.

\bibitem[Lehtinen {\em et~al.\/}(1999)Lehtinen, Bell \&
  Inan]{lehtinen1999monte}
{\sc \au{Lehtinen, N.~G.}, \au{Bell, T.~F.} \& \au{Inan, U.~S.}} \yr{1999}
  \at{{Monte Carlo} simulation of runaway {MeV} electron breakdown with
  application to red sprites and terrestrial gamma ray flashes}.  \jt{Journal
  of Geophysical Research: Space Physics}  \bvol{104}~(A11),
  \pg{24699--24712}.

\bibitem[Linder {\em et~al.\/}(2020)Linder, Fable, Jenko, Papp, Pautasso \&
  and]{Linder_2020}
{\sc \au{Linder, O.}, \au{Fable, E.}, \au{Jenko, F.}, \au{Papp, G.},
  \au{Pautasso, G.} \& \au{and}} \yr{2020}  \at{Self-consistent modeling of
  runaway electron generation in massive gas injection scenarios in {ASDEX}
  upgrade}.  \jt{Nuclear Fusion}  \bvol{60}~(9),  \pg{096031}.

\bibitem[Lvovskiy {\em et~al.\/}(2018)Lvovskiy, Paz-Soldan, Eidietis, Molin,
  Du, Giacomelli, Herfindal, Hollmann, Martinelli, Moyer, Nocente, Rigamonti,
  Shiraki, Tardocchi \& Thome]{Lvovskiy2018}
{\sc \au{Lvovskiy, A.}, \au{Paz-Soldan, C.}, \au{Eidietis, N.~W.}, \au{Molin,
  A.~Dal}, \au{Du, X.~D.}, \au{Giacomelli, L.}, \au{Herfindal, J.~L.},
  \au{Hollmann, E.~M.}, \au{Martinelli, L.}, \au{Moyer, R.~A.}, \au{Nocente,
  M.}, \au{Rigamonti, D.}, \au{Shiraki, D.}, \au{Tardocchi, M.} \& \au{Thome,
  K.~E.}} \yr{2018}  \at{The role of kinetic instabilities in formation of the
  runaway electron current after argon injection in {DIII}-{D}}.  \jt{Plasma
  Physics and Controlled Fusion}  \bvol{60}~(12),  \pg{124003}.

\bibitem[Mart\'{\i}n-Sol\'{\i}s {\em et~al.\/}(2015)Mart\'{\i}n-Sol\'{\i}s,
  Loarte \& Lehnen]{martinsolis1}
{\sc \au{Mart\'{\i}n-Sol\'{\i}s, J.~R.}, \au{Loarte, A.} \& \au{Lehnen, M.}}
  \yr{2015}  \at{Runaway electron dynamics in tokamak plasmas with high
  impurity content}.  \jt{Physics of Plasmas}  \bvol{22},  \pg{092512}.

\bibitem[Mart{\'\i}n-Sol{\'\i}s {\em et~al.\/}(2017)Mart{\'\i}n-Sol{\'\i}s,
  Loarte \& Lehnen]{martin2017formation}
{\sc \au{Mart{\'\i}n-Sol{\'\i}s, J.~R.}, \au{Loarte, A.} \& \au{Lehnen, M.}}
  \yr{2017}  \at{Formation and termination of runaway beams in {ITER}
  disruptions}.  \jt{Nuclear Fusion}  \bvol{57}~(6),  \pg{066025}.

\bibitem[Mart\'{\i}n-Sol\'{\i}s {\em et~al.\/}(2010)Mart\'{\i}n-Sol\'{\i}s,
  S\'anchez \& Esposito]{MartinSolis}
{\sc \au{Mart\'{\i}n-Sol\'{\i}s, J.~R.}, \au{S\'anchez, R.} \& \au{Esposito,
  B.}} \yr{2010}  \at{Experimental observation of increased threshold electric
  field for runaway generation due to synchrotron radiation losses in the {FTU}
  tokamak}.  \jt{Phys. Rev. Lett.}  \bvol{105},  \pg{185002}.

\bibitem[Matsuyama {\em et~al.\/}(2017)Matsuyama, Aiba \& Yagi]{Matsuyama_2017}
{\sc \au{Matsuyama, A.}, \au{Aiba, N.} \& \au{Yagi, M.}} \yr{2017}  \at{Reduced
  fluid simulation of runaway electron generation in the presence of resistive
  kink modes}.  \jt{Nuclear Fusion}  \bvol{57}~(6),  \pg{066038}.

\bibitem[McDevitt \& Tang(2019)]{McDevitt_2019}
{\sc \au{McDevitt, C.~J.} \& \au{Tang, X.-Z.}} \yr{2019}  \at{Runaway electron
  generation in axisymmetric tokamak geometry}.  \jt{{EPL} (Europhysics
  Letters)}  \bvol{127}~(4),  \pg{45001}.

\bibitem[Mlynar {\em et~al.\/}(2018)Mlynar, Ficker, Macusova, Markovic,
  Naydenkova, Papp, Urban, Vlainic, Vondracek, Weinzettl, Bogar, Bren,
  Carnevale, Casolari, Cerovsky, Farnik, Gobbin, Gospodarczyk, Hron, Kulhanek,
  Havlicek, Havranek, Imrisek, Jakubowski, Lamas, Linhart, Malinowski,
  Marcisovsky, Matveeva, Panek, Plyusnin, Rabinski, Svoboda, Svihra, Varju \&
  Zebrowski]{Mlynar_2018}
{\sc \au{Mlynar, J.}, \au{Ficker, O.}, \au{Macusova, E.}, \au{Markovic, T.},
  \au{Naydenkova, D.}, \au{Papp, G.}, \au{Urban, J.}, \au{Vlainic, M.},
  \au{Vondracek, P.}, \au{Weinzettl, V.}, \au{Bogar, O.}, \au{Bren, D.},
  \au{Carnevale, D.}, \au{Casolari, A.}, \au{Cerovsky, J.}, \au{Farnik, M.},
  \au{Gobbin, M.}, \au{Gospodarczyk, M.}, \au{Hron, M.}, \au{Kulhanek, P.},
  \au{Havlicek, J.}, \au{Havranek, A.}, \au{Imrisek, M.}, \au{Jakubowski, M.},
  \au{Lamas, N.}, \au{Linhart, V.}, \au{Malinowski, K.}, \au{Marcisovsky, M.},
  \au{Matveeva, E.}, \au{Panek, R.}, \au{Plyusnin, V.~V.}, \au{Rabinski, M.},
  \au{Svoboda, V.}, \au{Svihra, P.}, \au{Varju, J.} \& \au{Zebrowski, J.}}
  \yr{2018}  \at{Runaway electron experiments at {COMPASS} in support of the
  {EUROfusion} {ITER} physics research}.  \jt{Plasma Physics and Controlled
  Fusion}  \bvol{61}~(1),  \pg{014010}.

\bibitem[Myra \& Catto(1992)]{Myra_1992}
{\sc \au{Myra, J.~R.} \& \au{Catto, P.~J.}} \yr{1992}  \at{Effect of drifts on
  the diffusion of runaway electrons in tokamak stochastic magnetic fields}.
  \jt{Phys. Fluids B: Plasma Phys.}  \bvol{4}~(1),  \pg{176--186}.

\bibitem[Papp {\em et~al.\/}(2011)Papp, Drevlak, F\"ul\"op, Helander \&
  Pokol]{Papp_2011PPCF}
{\sc \au{Papp, G.}, \au{Drevlak, M.}, \au{F\"ul\"op, T.}, \au{Helander, P.} \&
  \au{Pokol, G.~I.}} \yr{2011}  \at{Runaway electron losses caused by resonant
  magnetic perturbations in {ITER}}.  \jt{Plasma Phys. Control. Fusion}
  \bvol{53}~(9),  \pg{095004}.

\bibitem[Papp {\em et~al.\/}(2015)Papp, Drevlak, Pokol \& Fülöp]{Papp_2015}
{\sc \au{Papp, G.}, \au{Drevlak, M.}, \au{Pokol, G.~I.} \& \au{Fülöp, T.}}
  \yr{2015}  \at{Energetic electron transport in the presence of magnetic
  perturbations in magnetically confined plasmas}.  \jt{J. Plasma Phys.}
  \bvol{81}~(5).

\bibitem[Papp {\em et~al.\/}(2013)Papp, Fülöp, Feh{\'{e}}r, de~Vries,
  Riccardo, Reux, Lehnen, Kiptily, Plyusnin \& and]{Papp_2013}
{\sc \au{Papp, G.}, \au{Fülöp, T.}, \au{Feh{\'{e}}r, T.}, \au{de~Vries,
  P.C.}, \au{Riccardo, V.}, \au{Reux, C.}, \au{Lehnen, M.}, \au{Kiptily, V.},
  \au{Plyusnin, V.V.} \& \au{and, B.~Alper}} \yr{2013}  \at{The effect of
  {ITER}-like wall on runaway electron generation in {JET}}.  \jt{Nuclear
  Fusion}  \bvol{53}~(12),  \pg{123017}.

\bibitem[Parail {\em et~al.\/}(2013)Parail, Albanese, Ambrosino, Artaud,
  Besseghir, Cavinato, Corrigan, Garcia, Garzotti, Gribov, Imbeaux, Koechl,
  Labate, Lister, Litaudon, Loarte, Maget, Mattei, McDonald, Nardon, Saibene,
  Sartori \& Urban]{Parail_2013}
{\sc \au{Parail, V.}, \au{Albanese, R.}, \au{Ambrosino, R.}, \au{Artaud,
  J.-F.}, \au{Besseghir, K.}, \au{Cavinato, M.}, \au{Corrigan, G.}, \au{Garcia,
  J.}, \au{Garzotti, L.}, \au{Gribov, Y.}, \au{Imbeaux, F.}, \au{Koechl, F.},
  \au{Labate, C.V.}, \au{Lister, J.}, \au{Litaudon, X.}, \au{Loarte, A.},
  \au{Maget, P.}, \au{Mattei, M.}, \au{McDonald, D.}, \au{Nardon, E.},
  \au{Saibene, G.}, \au{Sartori, R.} \& \au{Urban, J.}} \yr{2013}
  \at{Self-consistent simulation of plasma scenarios for {ITER} using a
  combination of {1.5D} transport codes and free-boundary equilibrium codes}.
  \jt{Nuclear Fusion}  \bvol{53}~(11),  \pg{113002}.

\bibitem[Paz-Soldan {\em et~al.\/}(2014)Paz-Soldan, Eidietis, Granetz,
  Hollmann, Moyer, Wesley, Zhang, Austin, Crocker, Wingen \& Zhu]{Paz-Soldan}
{\sc \au{Paz-Soldan, C.}, \au{Eidietis, N.~W.}, \au{Granetz, R.}, \au{Hollmann,
  E.~M.}, \au{Moyer, R.~A.}, \au{Wesley, J.~C.}, \au{Zhang, J.}, \au{Austin,
  M.~E.}, \au{Crocker, N.~A.}, \au{Wingen, A.} \& \au{Zhu, Y.}} \yr{2014}
  \at{Growth and decay of runaway electrons above the critical electric field
  under quiescent conditions}.  \jt{Physics of Plasmas}  \bvol{21}~(2),
  \pg{022514}.

\bibitem[Popovic {\em et~al.\/}(2016)Popovic, Esposito, Mart\'{\i}n-Sol\'{\i}s,
  Bin, Buratti, Carnevale, Causa, Gospodarczyk, Marocco, Ramogida \&
  Riva]{Popovic}
{\sc \au{Popovic, Z.}, \au{Esposito, B.}, \au{Mart\'{\i}n-Sol\'{\i}s, J.~R.},
  \au{Bin, W.}, \au{Buratti, P.}, \au{Carnevale, D.}, \au{Causa, F.},
  \au{Gospodarczyk, M.}, \au{Marocco, D.}, \au{Ramogida, G.} \& \au{Riva, M.}}
  \yr{2016}  \at{On the measurement of the threshold electric field for runaway
  electron generation in the {Frascati} tokamak upgrade}.  \jt{Physics of
  Plasmas}  \bvol{23}~(12),  \pg{122501}.

\bibitem[Rechester \& Rosenbluth(1978)]{rechester1978electron}
{\sc \au{Rechester, A.~B.} \& \au{Rosenbluth, M.~N.}} \yr{1978}  \at{Electron
  heat transport in a tokamak with destroyed magnetic surfaces}.  \jt{Physical
  Review Letters}  \bvol{40}~(1),  \pg{38}.

\bibitem[Rosenbluth \& Putvinski(1997)]{RosenbluthPutvinski1997}
{\sc \au{Rosenbluth, M.~N.} \& \au{Putvinski, S.~V.}} \yr{1997}  \at{Theory for
  avalanche of runaway electrons in tokamaks}.  \jt{Nuclear Fusion}  \bvol{37},
   \pg{1355--1362}.

\bibitem[Smith {\em et~al.\/}(2006)Smith, Helander, Eriksson, Anderson, Lisak
  \& Andersson]{smith2006runaway}
{\sc \au{Smith, H.}, \au{Helander, P.}, \au{Eriksson, L-G.}, \au{Anderson, D.},
  \au{Lisak, M.} \& \au{Andersson, F.}} \yr{2006}  \at{Runaway electrons and
  the evolution of the plasma current in tokamak disruptions}.  \jt{Physics of
  Plasmas}  \bvol{13}~(10),  \pg{102502}.

\bibitem[Spitzer \& H{\"a}rm(1953)]{spitzer1953transport}
{\sc \au{Spitzer, L.} \& \au{H{\"a}rm, R.}} \yr{1953}  \at{Transport phenomena
  in a completely ionized gas}.  \jt{Physical Review}  \bvol{89}~(5),
  \pg{977}.

\bibitem[Stahl {\em et~al.\/}(2015)Stahl, Hirvijoki, Decker, Embr\'eus \&
  F\"{u}l\"{o}p]{Stahl2015}
{\sc \au{Stahl, A.}, \au{Hirvijoki, E.}, \au{Decker, J.}, \au{Embr\'eus, O.} \&
  \au{F\"{u}l\"{o}p, T.}} \yr{2015}  \at{Effective critical electric field for
  runaway electron generation}.  \jt{Physical Review Letters}  \bvol{114},
  \pg{115002}.

\bibitem[Särkimäki {\em et~al.\/}(2020)Särkimäki, Embreus, Nardon,
  F\"ul\"op \& {JET Contributors}]{Sarkimaki_2020}
{\sc \au{Särkimäki, K.}, \au{Embreus, O.}, \au{Nardon, E.}, \au{F\"ul\"op,
  T.} \& \au{{JET Contributors}}} \yr{2020}  \at{Assessing energy dependence of
  the transport of relativistic electrons in perturbed magnetic fields with
  orbit-following simulations}.  \jt{Nuclear Fusion} .

\bibitem[Särkimäki {\em et~al.\/}(2016)Särkimäki, Hirvijoki, Decker, Varje
  \& Kurki-Suonio]{Sarkimaki_2016}
{\sc \au{Särkimäki, K.}, \au{Hirvijoki, E.}, \au{Decker, J.}, \au{Varje, J.}
  \& \au{Kurki-Suonio, T.}} \yr{2016}  \at{An advection{\textendash}diffusion
  model for cross{\textendash}field runaway electron transport in perturbed
  magnetic fields}.  \jt{Plasma Phys. Control. Fusion}  \bvol{58}~(12),
  \pg{125017}.

\bibitem[Vallhagen {\em et~al.\/}(2020)Vallhagen, Embreus, Pusztai, Hesslow \&
  Fülöp]{Vallhagen-JPP-20}
{\sc \au{Vallhagen, O.}, \au{Embreus, O.}, \au{Pusztai, I.}, \au{Hesslow, L.}
  \& \au{Fülöp, T.}} \yr{2020}  \at{Runaway dynamics in the {DT} phase of
  {ITER} operations in the presence of massive material injection}.
  \jt{Journal of Plasma Physics}  \bvol{86}~(4),  \pg{475860401}.

\bibitem[Varje {\em et~al.\/}(2019)Varje, Särkimäki, Kontula, Ollus,
  Kurki-Suonio, Snicker, Hirvijoki \& Äkäslompolo]{ascot5}
{\sc \au{Varje, J.}, \au{Särkimäki, K.}, \au{Kontula, J.}, \au{Ollus, P.},
  \au{Kurki-Suonio, T.}, \au{Snicker, A.}, \au{Hirvijoki, E.} \&
  \au{Äkäslompolo, S.}} \yr{2019} {High-performance orbit-following code
  ASCOT5 for Monte Carlo simulations in fusion plasmas}. {Submitted to Comp.
  Phys. Comm.},  \arxiv{arXiv: 1908.02482}.

\bibitem[Ward \& Wesson(1992)]{ward1992impurity}
{\sc \au{Ward, D.~J.} \& \au{Wesson, J.~A.}} \yr{1992}  \at{Impurity influx
  model of fast tokamak disruptions}.  \jt{Nuclear fusion}  \bvol{32}~(7),
  \pg{1117}.

\bibitem[Yoshino \& Tokuda(2000)]{Yoshino_2000}
{\sc \au{Yoshino, R.} \& \au{Tokuda, S.}} \yr{2000}  \at{Runaway electrons in
  magnetic turbulence and runaway current termination in tokamak discharges}.
  \jt{Nucl. Fusion}  \bvol{40}~(7),  \pg{1293--1309}.

\bibitem[Zeng {\em et~al.\/}(2017)Zeng, Chen, Dong, Koslowski, Liang, Zhang,
  Zhuang, Huang \& Gao]{Zeng2017}
{\sc \au{Zeng, L.}, \au{Chen, Z.Y.}, \au{Dong, Y.B.}, \au{Koslowski, H.R.},
  \au{Liang, Y.}, \au{Zhang, Y.P.}, \au{Zhuang, H.D.}, \au{Huang, D.W.} \&
  \au{Gao, X.}} \yr{2017}  \at{Runaway electron generation during disruptions
  in the {J}-{TEXT} tokamak}.  \jt{Nuclear Fusion}  \bvol{57}~(4),
  \pg{046001}.

\bibitem[Zeng {\em et~al.\/}(2013)Zeng, Koslowski, Liang, Lvovskiy, Lehnen,
  Nicolai, Pearson, Rack, Jaegers, Finken, Wongrach, Xu \& the
  TEXTOR~team]{Zeng2013}
{\sc \au{Zeng, L.}, \au{Koslowski, H.~R.}, \au{Liang, Y.}, \au{Lvovskiy, A.},
  \au{Lehnen, M.}, \au{Nicolai, D.}, \au{Pearson, J.}, \au{Rack, M.},
  \au{Jaegers, H.}, \au{Finken, K.~H.}, \au{Wongrach, K.}, \au{Xu, Y.} \&
  \au{the TEXTOR~team}} \yr{2013}  \at{Experimental observation of a
  magnetic-turbulence threshold for runaway-electron generation in the {TEXTOR}
  tokamak}.  \jt{Phys. Rev. Lett.}  \bvol{110},  \pg{235003}.

\end{thebibliography}

\end{document}